\documentclass[manuscript,screen]{acmart}

\AtBeginDocument{%
  \providecommand\BibTeX{{%
    \normalfont B\kern-0.5em{\scshape i\kern-0.25em b}\kern-0.8em\Te\X}}}
\usepackage{xcolor}
\usepackage{hyperref}
\begin{document}

\title{An Empirical Study On Correlation between Readme Content and Project Popularity
}

\author{Akhila Sri Manasa Venigalla}
\email{cs19d504@iittp.ac.in}
\author{Sridhar Chimalakonda}
\email{ch@iittp.ac.in}
\affiliation{%
  \institution{\\Indian Institute of Technology Tirupati}
  \city{Tirupati}
  \country{India}
}

\renewcommand{\shortauthors}{Venigalla and Chimalakonda}

\begin{abstract}
Readme in GitHub repositories serves as a preliminary source of information, and thus helps developers in understanding about the projects, for reuse or extension. Different types of contextual and structural content, which we refer to as \textit{categories} of the content and \textit{features} in the content respectively, are present in readme files, and could determine the extent of comprehension about project. Consequently, the structural and contextual aspects of the content could impact the project popularity. Studying the correlation between the content and project popularity could help in focusing on the aspects that could improve popularity, while designing the readme files. However, existing studies explore the categories of \textit{content} and types of \textit{features} in readme files, and do not explore their usefulness towards project popularity. Hence, we present an empirical study to understand correlation between readme file content and project popularity. We perform the study on 1950 readme files of public GitHub projects, spanning across ten programming languages, and observe that readme files in majority of the popular projects are well organised using lists and images, and comprise links to external sources. Also, repositories with readme files containing contribution guidelines and references were observed to be associated with higher popularity.

\end{abstract}


\maketitle

 \section{Introduction}
 The number of open source projects available on platforms such as GitHub are increasing day-by-day. GitHub alone hosts more than 200 million repositories, involving more than 83 million developers, as of May, 2022\footnote{\url{https://github.com/about}}. Developers across the world contribute to these projects for their development, identifying and resolving bugs, introducing new features and so on \cite{bao2019large, milewicz2019characterizing}. Parts of the projects, such as certain code snippets or the projects as a whole, are reused for development of newer projects on GitHub \cite{gharehyazie2019cross}. Readme is one of the important artifacts of the repository, that is interacted for contribution or reuse, while also supporting further analysis on the repositories \cite{prana2019categorizing, zhang2017detecting}.
 
 Readme files contain variety of information that includes instructions on using the project, the features of the repository, contribution guidelines, release information and so on \cite{prana2019categorizing, cogo2021empirical}, {which is used both by developers and researchers \cite{prana2021including, liu2021reproducibility, mitchell2022}. The information in readme file largely impacts the contribution or usage of a repository \cite{prana2021including}. Studies have also revealed that unclear and insufficient readme files have lead to poor developer onboarding and reduced contributions from developers \cite{prana2021including}. Studies on developer discussions, such as gitter platform, have also indicated that developers consult documentation in the form of readme files for issue resolution and better project comprehension \cite{ehsan2020empirical}.} This in turn impacts the popularity and progress of the repository as introducing new features, resolving bugs and so on depends on contributors of the repository \cite{tsay2014let}.  Prana et al. have categorised the content in readme files into seven categories to facilitate better and easier discovery of information present and consequently update the file in cases of missing categories  \cite{prana2019categorizing}. The content in readme files has been analysed in a recent study to understand the factors that impact popularity of academic AI repositories \cite{fan2021makes}. Patterns in readme files of java-based repositories on GitHub were identified, and their association with project popularity has also been analysed \cite{liu2022readme}. 
 While the consistent changes in readme files were observed to be correlated to project popularity \cite{aggarwal2014co}, these studies further indicate that content in readme files could have an impact on the popularity of the repository, along with the frequency of changes to readme files.

 Analysing the content that differs across popular and non-popular readme files could provide insights on useful and important content that should be present in readme files for better project popularity. This analysis could also lead to understanding the common practices of logging readme files among the popular projects.
 However, there exist very few empirical studies in the literature that study the correlation between content in readme files and properties of project. While Fan et al. \cite{fan2021makes} have considered readme file content in assessing the popularity of projects, this study was specific to academic AI repositories. Also, the study conducted by Liu et al. \cite{liu2022readme} analyzes the extent of conformance of readme content to official GitHub guidelines, and compares the presence with number of stars associated with the project. However, this study is specific to projects in java programming language. Hence, in this paper, we plan to perform an empirical study on the \textbf{correlation of readme content with project popularity}, irrespective of the type of repositories. We include fork count, watcher count and pullrequest count, along with star count in determining popularity of the project, as indicated in the literature \cite{borges2018s,chowdhury2016characterizing,aggarwal2014co}. 
 
 We perform an empirical study to identify \begin{itemize}
     \item \textbf{RQ1} -  Correlation between features (structural content) in the readme files and the project popularity.
     \item \textbf{RQ2} - Correlation between categories (contextual content) of content present in readme files and the project popularity
     \item \textbf{RQ3} - Features and categories of readme content present in popular repositories, which could be important for better comprehension of the repositories.
 \end{itemize} We determine popularity of the repositories by considering stars, forks, watchers and pull requests, which were observed to be correlated with project popularity in the literature \cite{dabbish2012social,aggarwal2014co,lee2013github,fan2021makes}. 
 The RQ1 and RQ2 could provide insights on the relation between project popularity and each readme category and feature, while RQ3 could provide insights on the important features and categories that are present/absent in popular versus non-popular repositories. RQ3 could thus drive us towards a list of important readme categories and features prevalent in popular repositories.
 
To answer the above research questions, we study the content in readme files of around 2000 GitHub repositories spanning across top 10 popular programming languages on GitHub, listed in Table \ref{tab:language}. The programming languages were considered based on their order of popularity on GitHub, as we carryout the analysis on GitHub repositories. This order of popularity was observed to differ with the standard TIOBE Index (presented in Table \ref{tab:language}). However, while order of popularity differed, 7 out of the 10 programming languages considered had the TIOBE Index value less than 10, indicating that majority (70\%) of the programming languages considered are widely popular, thus suggesting generalisation of results of the study.
 \begin{table}[]
\begin{tabular}{|c|c|c|}
\hline
\textbf{Language} & \textbf{\begin{tabular}[c]{@{}c@{}}GH Popularity\\ Rank\end{tabular}} & \textbf{TIOBE Index} \\ \hline
C                 & 10                                                                    & 2                    \\ \hline
C++               & 6                                                                     & 4                    \\ \hline
C\#               & 9                                                                     & 5                    \\ \hline
Go                & 5                                                                     & 13                   \\ \hline
Java              & 3                                                                     & 3                    \\ \hline
Javascript        & 2                                                                     & 7                    \\ \hline
PHP               & 8                                                                     & 8                    \\ \hline
Python            & 1                                                                     & 1                    \\ \hline
Ruby              & 7                                                                     & 16                   \\ \hline
Typescript        & 4                                                                     & 40                   \\ \hline
\end{tabular}
\caption{The top 10 popular programming languages and their corresponding ranking in terms of popularity}
\label{tab:language}
\end{table}
We devise an approach to automatically calculate the feature count present in readme files for each of the features. {The content in readme files could be categorized using Latent Dirichlet Allocation (LDA) topic modelling approach \cite{blei2003latent}.
LDA approach can be assigned to classify the content into eight categories based on the seven categories mentioned in \cite{prana2019categorizing}, and the eighth category corresponding to `Others', which includes content that does not belong to any of the seven categories. The keywords in each of the categories should be analysed and labelled accordingly, with the seven categories. Sections in readme files can be identified based on the header notations, and the content in each section can then be analysed for the presence of keywords corresponding to the eight categories (seven + others). The content can be labelled with the category based on keywords, using machine learning classifiers such as SVM \cite{hearst1998svm}, Random Forest \cite{breiman2001random} and so on. A similar approach is presented in the readme classifier provided by Prana et al. in \cite{prana2019categorizing} to categorize the content present in readme files, and hence was used to identify categories in readme files of the selected repositories.}  We then compare the distribution of these features and categories with project popularity, towards answering the research questions.

\textbf{\textit{We define features as the structural components involved in presenting the readme files, such as lists, images, external links, code components and so on.
}}

\textbf{\textit{We define categories of content based on the type of content present in the readme files, such as installation instructions, contribution guidelines and so on.
}}

The main contributions of the paper are:
\begin{itemize}
    \item An empirical analysis of features and categories in 1950 readme files corresponding to 2000 GitHub repositories across top 10 popular programming languages on GitHub.
    \item A list of features and categories in readme files, and their correlation with the popularity of repositories. The results of the study are presented in Section \ref{res}.
    \item A ranked list of important features and categories that could potentially support better project comprehension, assuming popular repositories to have better comprehension.
\end{itemize}

The rest of the paper is organised as follows -  Section \ref{rel} discusses some of the works in the literature that analysed the popularity of repositories based on various factors. Section \ref{design} describes the methodology followed in the study and Section \ref{res} discusses the results of the study. The threats to validity in the study are presented in Section \ref{ttv}.	The implications of the study to researchers and practitioners are discussed in Section \ref{disc}, followed by conclusion in Section \ref{concl}.

 



\section{Related Work}
\label{rel}
Studies in the literature have analysed project characteristics of open source projects, based on their correlation with artifacts such as issues \cite{kavaler2017perceived}, pull requests \cite{aggarwal2014co} and forks \cite{zhu2014patterns} associated with the project, and based on the content in readme files such as badges \cite{trockman2018adding}, keywords \cite{overney2020hanging}, and textual categories \cite{fan2021makes}. We group the most relevant studies in literature that aim towards understanding project characteristics on the basis of artifacts analysed, into two categories - (i) corresponding to  correlation with artifact features and (ii) corresponding to readme files.

\subsection{Studies corresponding to correlation of artifact features and project characteristics}
There exists works in the literature that explore the relationship between various characteristics of the repositories and the features of different artifacts in the repositories. 

For example, the technical complexity of language in issue discussions has been analysed, which revealed that discussions inline with the project-specific language contribute to faster issue resolution rates \cite{kavaler2017perceived}. Other characteristics such as determining pull request acceptance based on contributors' profile \cite{yu2015wait}, issue resolution based on type of maintenance activity in the project and so on were explored in the literature \cite{murgia2014influence}.

Popularity is among the most commonly considered characteristics of the project. 
Borges et al. have analysed correlation between different project characteristics such as fork count, commit count, repository age and so on, and the project popularity of top starred 2500 repositories on GitHub, by considering star count as popularity metric \cite{borges2016understanding}. 
A considerable difference in the values of these characteristics was observed in relation to the star-count of the repositories \cite{borges2016understanding}. These features were further used in generating prediction models to predict the number of stars and ranking of GitHub repositories \cite{borges2016predicting}. 
Zhu et al. defined popularity of the projects based on the projects' fork count and analysed the variation of fork counts of projects based on the patterns of folder structure in projects \cite{zhu2014patterns}. The folder structure and types were analysed for 140K GitHub projects, where the standard folders such as those corresponding to testing and usage examples were observed to contribute to increased fork counts \cite{zhu2014patterns}. 

The updates made and their frequency to documentation related files in different types of projects were studied to assess if they contribute to popularity of the projects, with number of stars, forks and pull requests being considered as dimensions of popularity \cite{aggarwal2014co}. This study revealed that consistent updates to documentation related artifacts contribute to increased popularity of the projects \cite{aggarwal2014co}.
Twenty features corresponding to source code, replicability and documentation of the projects have been studied to understand their impact on the star count of the academic AI repositories \cite{fan2021makes}. The relation between different combinations of content in java readme files and star count of the corresponding repositories has been analysed, to find the most frequently used patterns in readme files of popular java repositories \cite{liu2022readme}.

\subsection{Studies corresponding to readme files}
Readme files are considered as a preliminary source to understand about a repository, both by researchers and developers. Readme files have been used to assess and understand various properties of repositories in the literature \cite{trockman2018adding, fan2021makes, liu2022readme}.

Zhang et al. have analysed the content in readme files to detect similarities among the repositories \cite{zhang2017detecting}. 
The quality of around 290K GitHub repositories from project maintainers' perspective is assessed by obtaining a degree of correlation of content corresponding to badges in readme files with project quality \cite{trockman2018adding}. 
In an attempt to understand types of content present in Readme files, Prana et al. have analysed the readme content of 393 GitHub repositories. The content was then classified into seven categories based on the information being discussed in each section of the readme files \cite{prana2019categorizing}. This categorisation was perceived to be useful to the developers and a classifier has been designed to automatically detect such categories in the readme files, to ease the process of locating information in readme files \cite{prana2019categorizing}.

The content in readme files of 77 million GitHub repositories presented by GHTorrent has been analysed  for the presence of links or keywords pointing to "donations", to identify donation platforms \cite{overney2020hanging}. 
The quality of readme content has been analysed to understand the quality of software documentation in projects \cite{treude2020beyond}.
Ten dimensions that define quality of a document have been presented and their presence is checked against readme files of 159 GitHub projects corresponding to R language manually \cite{treude2020beyond}.
Towards understanding the factors that influence adoption of CI tools to Github repositories, the readme content is processed, to extract repository badges, that provide insights on the nature of the projects, which could consequently help in identifying features of the projects \cite{lamba2020heard}.

The readme files, along with the source code of about 1.1K GitHub repositories in the context of AI have been analysed to identify the features in the readme content that contribute to popularity of the repositories \cite{fan2021makes}. About 10 features corresponding to the documentation were presented and were analysed for contributions to popularity of the repositories \cite{fan2021makes}.
The content in readme files of 129 GitHub projects, corresponding to COVID-19, was analysed to identify the category of projects, in an attempt to identify the relation between different types of bugs and the project categories, in projects dealing with COVID-19 \cite{rahman2021empirical}.
Readme files of 14.9K java repositories on GitHub have been clustered based on the content present in them, to assess their conformance to GitHub official guidelines, to detect common patterns, and to identify the relationship between star count and the presence of specific groups of content in the repositories \cite{liu2022readme}.



Features of various artifacts such as frequency of updates, number of contributors and so on were analysed for their correlation with popularity. Though readme files are one of the important artifacts of the project, there exist very few studies that assess contents in the readme files, especially towards their contribution to project popularity. While Fan et al. and Liu et al. explore the impact of different features of readme files on the project popularity, these studies focuses only on academic ai repositories and java-based repositories, and consider only star count as a metric for project popularity \cite{fan2021makes, liu2022readme}. 

Hence, through this study, we wish to extend the scope, to include GitHub repositories irrespective of their domain and programming language, to consider both structural and contextual characteristics of the readme content. We also aim to identify if there exists a difference across programming languages with respect to the categories and features present and their relation with project popularity, which could be determined based on the number of stars, forks, watchers and pull requests.

\section{Study Design}
\label{design}

The main aim of this paper is to assess the relation between project popularity and the features and the types of content in readme files. 
We envision that understanding the features and categories that strongly relate to project popularity could provide insights on specific type of content and features that could be considered by project developers and contributors for better project popularity. \textit{Assessing the presence of correlation between the features or categories and the project popularity could indicate the level of impact of these aspects on project popularity.}

To this end, we devise our study based on the Goal Question Metric approach \cite{basili1994goal}, which was used for conducting empirical studies in the literature that compare existing approaches or artifacts \cite{chen2020empirical}, presented in Table \ref{tab:gqm}, to explore the following research questions:


\textbf{RQ1: Does presence of a specific feature relate to the popularity of the repository?}

We process popular and non-popular repositories to extract the features present in the repository and then analyse the results using statistical hypothesis tests that support numerical values, to understand if each of the features have a significant relation with the popularity of the repository.
It has been observed that popular projects tend to have references with links to external sources and other GitHub repositories in their readme files, which were observed to be minimal among the non-popular repositories. The popular repositories included images in the readme files, which could provide more details about the projects for users and contributors.

\textbf{RQ2: Does presence of a specific category relate to the popularity of the repository?}
The popular and non-popular repositories were analysed for presence of categories in the readme files. We applied statistical significance tests, relevant to non-numerical values, to understand the significance of each category in contributing to popularity of the projects. 
Popular projects comprised more details in the readme files with respect to the contribution guidelines and references, which were hardly present in the non-popular projects. However, details about how to use a repository were present in readme files of majority of the repositories, irrespective of their popularity.
\begin{table*}[]
\begin{tabular}{llll}
\hline
\textbf{Goal}       &                           & \textbf{Question}                                & \textbf{Metric}                    \\
\hline

\textbf{Purpose:}   & Understanding             & \textbf{RQ1:} Does presence of a specific feature         & wilcoxon's rank sum test  \\
\textbf{Issue: }    & the impact                & impact the popularity of the repository?         &  p-value, cliff's delta                      \\
\textbf{Object:}    & of features \& categories & \textbf{RQ2:} Does presence of a specific category        & fisher's exact test p-value        \\
\textbf{Viewpoint:} & on project popularity     & impact the popularity of the repository?         &                                    \\
\textbf{Context:}   & in GitHub projects        & \textbf{RQ3:} Which features and categories  & gini importance score,             \\
           &                           & determine the popularity                     & permutation-based 
           \\
           &                           &  of the repository?                    & importance score
\\
\hline           
\end{tabular}
\caption{Summary of Goal Question Metric}
\label{tab:gqm}
\end{table*}

\textbf{RQ3: Which features and categories vary among the popular and non-popular repositories?}
Presence of a feature, or a category in the popular repositories and absence of the same in non-popular repositories, or viceversa, could indicate that the specific feature or category is capable of determining the popularity of the project. We build a random forest classifier to identify the important features and categories. 
To reduce the confusion and increase performance of the random forest classifier, features are analysed for any possible correlations among them, and one of the correlated features from each pair of correlated features was dropped while building the random forest classifier. This analysis of important features and categories revealed that presence of external links, links to other GitHub repositories, references and license information strongly differentiate the popular repositories from the non-popular ones.
\begin{figure*}
    \centering
    \includegraphics[width = \linewidth]{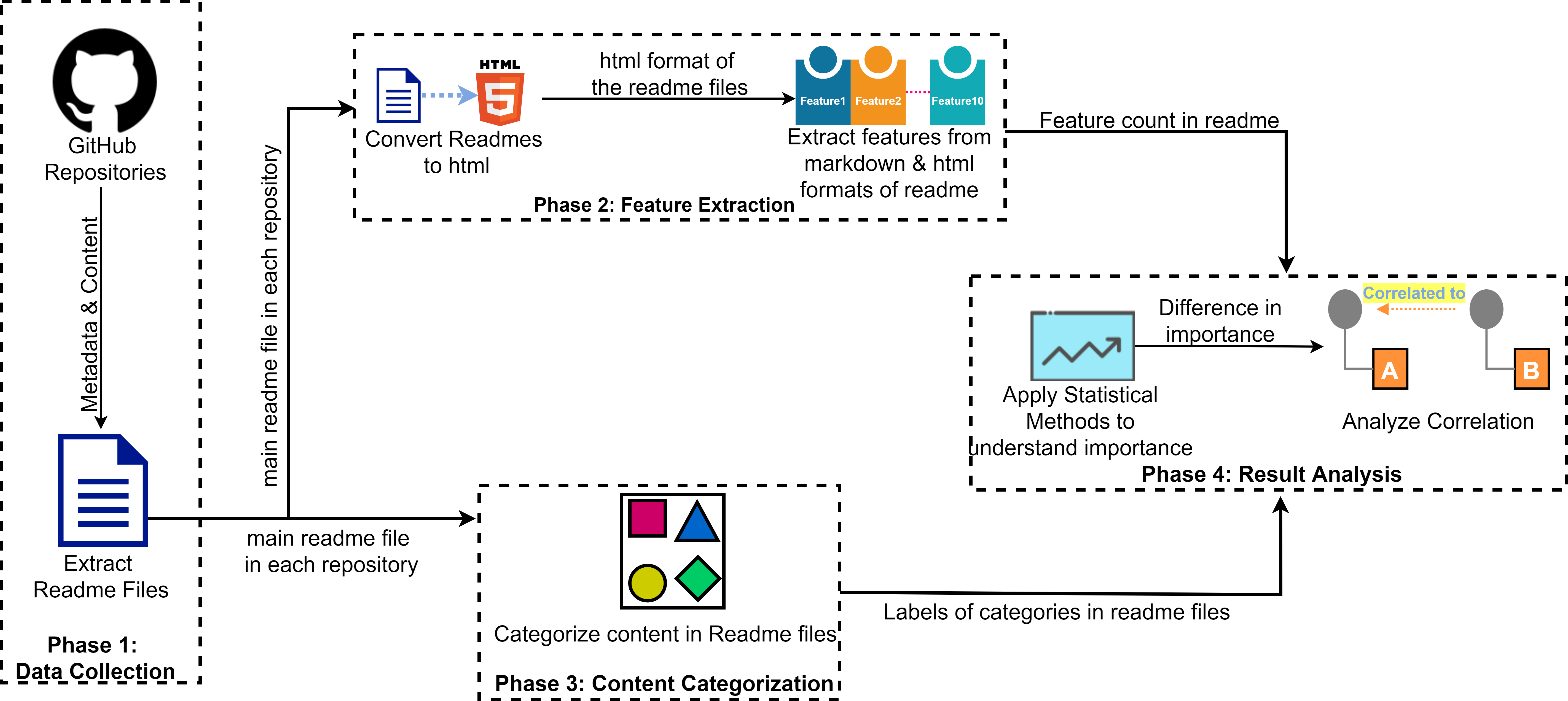}
    \caption{Methodology of the Empirical Study}
    \label{fig:meth}
\end{figure*}

We follow the four phase methodology presented in Figure \ref{fig:meth} that deals with data collection, data processing and result analysis at a higher level, similar to the empirical studies in literature that assess crowd sourced code and open-source projects \cite{chen2020empirical, verdi2020empirical, DBLP:conf/msr/CiniselliCPPPB21, DBLP:conf/msr/CandidoHAD21}, or the correlation between readme file content and popularity \cite{liu2022readme, fan2021makes}. For this empirical study, the data collection process deals with extracting the readme files of GitHub repositories (phase 1), while the data processing phase deals with feature extraction (phase 2) from, and categorization (phase 3) of, content in readme files and the result analysis phase deals with analysing the correlation between project popularity and features and categories using statistical approaches (phase 4). We elaborate each of the four phases below:


\subsection{Phase 1: Data Collection}
{Studies in the literature with similar goals of assessing correlation of presence of certain artifacts have analysed open source repositories ranging from about 100 to 77 million. The closest relevant works in the literature, that we consider as a basis for features \cite{fan2021makes} and categories \cite{prana2019categorizing}, consider 1.1K and 393 repositories respectively. Another work that analyses project characteristics based on star count \cite{borges2016understanding} considered 2.5K repositories. Similar to these existing studies, we collected 2000 GitHub repositories, which could be a sufficient sample of data, to assess the correlation between readme content and popularity of repositories. 
These repositories span across the top 10 most popular programming languages on GitHub\footnote{\url{https://madnight.github.io/githut/\#/pull_requests/2021/4}}.  200 repositories were collected from each of the 10 programming languages, and the metadata corresponding to number of stars, forks, watchers and pull requests has been extracted. }

{As an initial filtering criteria to support selection of repositories, we considered top starred 100 repositories and 100 repositories randomly, with star-count in the range of 0 to 100, from each programming language, to create an initial datadump of 2000 repositories.
The bottom most repository of the top starred 100 repositories across all the languages had the stargazer count as 5153\footnote{\url{https://github.com/rails-api/active_model_serializers}}, as of 10th December 2021. To ensure significant difference between the popular and non-popular repositories, we make an initial assumption that repositories with star count less than 100 could be considered as non-popular when compared to the popular repositories with at least 5153 stars. This also satisfies the assumptions in literature which considered repositories with at least 1000 stars to be popular \cite{jebnoun2020scent}. Thus, we consider star-count as an initial filter to identify popular and non-popular repositories, as star count is one of the prominent features to determine popularity \cite{liu2022readme,fan2021makes}. However, as popularity metric could also include other metrics of the project \cite{aggarwal2014co,dabbish2012social}, we later define the popularity metric to consider stars, forks, watchers and pull requests. We use this metric to assign popularity ranks to the repositories, by considering all 200 repositories across each of the programming languages. Thus, we consider star count as a prominent feature to determine popularity, but, we do not limit the popularity rank assignment solely to star count. 
}

For a well-distributed set of non-popular repositories, we extracted the metadata of 100 repositories randomly, with at least one repository with stargazer count in the range of 1 to 100 from each of the programming languages. 
While the popular repositories were considered based on the extreme values of stars (top 100 most starred repositories), the extreme values of stars (top 100 least starred repositories) could not be considered for non-popular repositories. This is because, majority of the least starred repositories had a star count of 0, thus not facilitating well-distributed nature of the dataset.
As a result, a list of 1000 popular and 1000 non-popular repositories across the ten programming languages have been extracted. This extraction took about 3 working days, due to the range limitations of GitHub API.

\begin{table*}[]
\begin{tabular}{|l|l|l|l|}
\hline
\textbf{Feature {[}f(x){]}} & \textbf{Feature description}             & \textbf{Identification {[}x{]} }                                       & \textbf{Source format} \\ \hline
List               & Number of lists                 & \textless{}ol\textgreater \& \textless{}ul\textgreater{}      & html          \\ \hline
Image              & Number of images                & \textless{}img\textgreater{}                                  & html          \\ \hline
Animated Image     & Number of animated images       & .gif                                                          & html          \\ \hline
Video              & Number of video links           & “video”, “youtu.be” in url sentence                           & html          \\ \hline
Table              & Number of tables                & \textless{}center\textgreater{}                               & html          \\ \hline
GitHub             & Number of github links          & “github.com” in url                                           & html          \\ \hline
Links              & Number of external links        & \textless{}a\textgreater and “url”                            & html          \\ \hline
Code               & Number of code blocks           & Half the number of 3 backticks                                & markdown      \\ \hline
Inline Code        & Number of inline code elements  & \textless{}code\textgreater \& \textless{}/code\textgreater{} & html          \\ \hline
Project            & Number of links to project page & "project" in "url"                                            & html          \\ \hline
\end{tabular}
\caption{Features considered for analysis of readme files}
\label{tab:features}
\end{table*}

We cloned all the 2000 repositories and processed these repositories to extract the readme files of interest across the repositories. The files with names `readme' and extensions - `.txt', `.md', `.rdoc', `.html', `.asciidoc', `.adoc', `.rst' and `.markdown' have been extracted. We observed that some of the repositories have multiple files with the name `readme' in the sub-folders of the projects, while some repositories do not comprise any files with the name `readme'. We limit the focus of this study to the readme files on the starting page of the GitHub repository, as they first appear while visiting a repository, and hence consider only those readme files in the root folder of each repository. Thus, we ensure that projects having readme content on the starting page of the GitHub repository are analysed. This resulted in a set of 1950 readme files that were analysed. All the readme files were renamed after their parent repository for better comprehension of the results.

\subsection{Phase 2: Feature extraction}
We explore different types of components present in the readme files. Fan et al. have presented 10 features in the documentation dimension in their study on popularity of academic AI repositories \cite{fan2021makes}. Readme files were considered for analysis in the study by Fan et al. to obtain insights on the documentation dimension. As we also aim to understand the impact of components present in readme files on popularity of the repository, similar to the study by Fan et al. \cite{fan2021makes}, we considered 9 out of the 10 features presented in documentation dimension by Fan et al in this study. We did not consider `has-license' feature presented in \cite{fan2021makes}, as this feature deals with presence of license information in the repository, but does not evaluate if the `license' component is present specifically in the readme file. 
We performed a manual walkthrough of readme files of 10 repositories, corresponding to languages being considered, on GitHub, to identify the presence of any other components, other than the 9 features being considered. We observed that readme files might contain links to external sources which need not be other GitHub repositories or those specific to the project. Thus, we add the tenth feature, `external links' to the set of features, and extract the 10 features, listed in Table \ref{tab:features}, from each of the repositories.

Considering the ease of extracting features presented in Table \ref{tab:features}, such as \textit{list}, \textit{tables}, and so on from html files, we convert the readme files written in markdown format to html using the markdown parser provided by python, the \textit{mistune}\footnote{\url{https://mistune.readthedocs.io/en/latest/}} package. 
The features \textit{links}, \textit{images}, \textit{animated images}, \textit{video links}, \textit{tables}, \textit{github links}, \textit{project links}, \textit{inline code} and \textit{external links} were extracted from the html version of the files, while the feature \textit{code} was extracted from the markdown version of the files.
Each feature, f(x) is extracted based on the presence of identification, x, and the resultant value, count\_feature\_fx, for each feature is calculated as follows:
\newline
\newline
\fbox{\begin{minipage}{46em}

for each\_line in html file, if “x” in each\_line, count\_feature\_fx = count\_feature\_fx + 1    
\end{minipage}}

i.e.,
\begin{gather*}
count_{x} \ =\ {\textstyle \sum _{lines}} instance_{x}
\end{gather*}
where, count\textsubscript{x} refers to the number of occurrences of a specific feature f(x) and instance\textsubscript{x} refers to the presence of the corresponding identification, x, in the corresponding source files.

\begin{table}[]
\begin{tabular}{|l|l|}
\hline
\textbf{Category}     & \textbf{Content Examples}                     \\ \hline
What         & Functionalities of the project                \\ \hline
Why          & Purpose of the project                        \\ \hline
How          & Installation instructions                     \\ \hline
When         & Release information and project status        \\ \hline
Who          & Contributors and project team                 \\ \hline
References   & API documentation \& other useful information \\ \hline
Contribution & Contribution guidelines                       \\ \hline
\end{tabular}
\caption{Categories considered for analysis of readme files}
\label{tab:categories}
\end{table}

\subsection{Phase 3: Content Categorization}
{
In order to categorize content in the readme files, we consider seven categories, as presented in \cite{prana2019categorizing} by Prana et al. and the GitHub guidelines\footnote{\url{https://guides.github.com/features/wikis/}} to create readme files \cite{liu2022readme}. In the study \cite{prana2019categorizing}, Prana et al., have proposed seven classes to categorize the content present in readme files of GitHub repositories. Since, the current study also deals with readme files in GitHub projects, we see that the categories presented by Prana et al. \cite{prana2019categorizing}, listed in Table \ref{tab:categories}, are relevant to the study. These categories were also observed to be inline with \textit{GitHub Recommended Sections} for readme files, and were integrated with the seven categories. The \textit{what} category corresponds to \textit{project description} section, \textit{how} category to \textit{installation and usage} section, \textit{contribution} category to \textit{contributing} section, and the \textit{who} category to \textit{credits} section of the recommended sections. Almost all the GitHub recommended guidelines for readme files were thus observed to be a subset of the categories presented in \cite{prana2019categorizing}, facilitating us to consider both the recommended guidelines as well as the categories proposed by Prana et al., by identifying the seven categories listed in Table \ref{tab:categories} in readme files. 
}

{The readme files extracted from each of the repositories are processed, and the section headers and corresponding content in the files are extracted. This extracted data is processed to remove stop words and generate tokens. Each of the tokens of a specific section in the readme file are analysed and mapped to one of the seven categories if they are related to one of the categories. This mapping is done with the help of a pre-trained SVM-based classifier, trained on manually annotated dataset as a part of READMEClassifier \cite{prana2019categorizing}. This classifier helps in identifying the types of categories present in the readme files, and also supports labelling of a specific section in the readme file with the corresponding category. With the readme files as input, a list of categories present in each of the files, based on the tokens in the files were obtained as output. Also, the categories of the corresponding sections for each readme file were presented as a result of classification. This data is further processed to include the boolean values `0' and `1' to each of the categories, to indicate presence or absence of a category in the readme file.
}

\subsection{Phase 4: Result Analysis}
The features and categories present in each of the repositories are compared to the popularity of the repositories to identify possible differences in the correlation of features and categories with respect to popularity. To facilitate this analysis, we consider popular and non-popular repositories to be two groups of data.
While stars were considered as the basis to extract repositories during data collection, works in the literature have explored varied approaches towards determining the popularity of repositories. To ensure credibility of repositories in popular and non-popular groups, we consider a two level popularity ranking mechanism as in \cite{jebnoun2020scent}, where the first level, in Phase 1, considers only stars and the second level, in this phase (Phase 4), includes other dimensions. 
Inline with the existing literature, we use the four dimensions - stars, forks, watchers and pull requests to assign popularity ranks to the repositories \cite{borges2018s,chowdhury2016characterizing,aggarwal2014co}. We define the popularity metric as the sum of these four dimensions, as follows:

\begin{verbatim}
    popularity = star_rank + fork_rank + watcher_rank + pullrequest_rank
    star_rank : position of repository in ascending order of star_count
    fork_rank : position of repository in ascending order of fork_count
    watcher_rank : position of repository in ascending order of watcher_count
    pullrequest_rank : position of repository in ascending order of pullrequest_count
\end{verbatim}

The star\_rank, fork\_rank, watcher\_rank and pullrequest\_rank  are calculated for all the repositories in each language in descending order. The repository with largest value is ranked with a larger value than others, with respect to each dimension. The ranks for all the four dimensions are added to generate popularity value. The popularity value is then compared across all the repositories corresponding to a specific programming language, and the resultant popularity\_rank is calculated similar to the dimension ranks, where the repository with largest popularity value is ranked with the highest value. Thus, the repository with larger popularity rank implies better popularity. For example, in the repository list of C programming language, the popularity rank of the repository \textit{netdata}\footnote{\url{https://github.com/netdata/netdata}} is 198, with star\_rank, fork\_rank, watcher\_rank and pullrequest\_rank being 197, 184, 197 and 193 respectively. The top ranked 100 repositories in terms of popularity are considered as popular repositories, while the rest were considered as non-popular repositories. While the order of popularity ranks varied slightly when compared to the star count, we observed that the top 100 repositories with respect to the popularity rank belonged to the top starred 100 repositories in each programming language. Thus, for each programming language, we arrive at a list of popular and non-popular repositories.

\subsubsection{Feature Analysis}
The count of presence of each feature for all the repositories has been extracted.
We then apply the Wilcoxon's rank sum test to obtain insights on whether there exists a significant difference in this count among popular and non-popular repository groups in each programming language.


\begin{table*}[]
\begin{tabular}{|l|cc|cccl|}
\hline
\multicolumn{1}{|c|}{\textbf{Feature}} & \multicolumn{2}{c|}{\textbf{Wilcoxon's p-value}}                                                                                                                                   & \multicolumn{4}{c|}{\textbf{Effect Size based on Cliff's delta}}                                                                                                                                                                                                                                                                             \\ \hline
\multicolumn{1}{|c|}{\textbf{}}        & \multicolumn{1}{c|}{\textbf{\textless{}0.05}}                                                      & \textbf{\textgreater{}0.05}                                                   & \multicolumn{1}{c|}{\textbf{Negligible}}                                                                    & \multicolumn{1}{c|}{\textbf{Small}}                                                              & \multicolumn{1}{c|}{\textbf{Medium}}     & \multicolumn{1}{c|}{\textbf{Large}}                                                                    \\ \hline
\textbf{List}                          & \multicolumn{1}{c|}{\begin{tabular}[c]{@{}c@{}}R, Py, Js, J,\\ C\#, Go, C, C++\end{tabular}}       & T, P                                                                          & \multicolumn{1}{c|}{T}                                                                             & \multicolumn{1}{c|}{P}                                                                       & \multicolumn{1}{c|}{R, C\#, C, C++} & \multicolumn{1}{c|}{Py, Js, J, Go}                                                                 \\ \hline
\textbf{Image}                         & \multicolumn{1}{c|}{\begin{tabular}[c]{@{}c@{}}R, T, Py, P, Js, J,\\ C\#, Go, C, C++\end{tabular}} & \multicolumn{1}{l|}{}                                                         & \multicolumn{1}{l|}{}                                                                              & \multicolumn{1}{c|}{T}                                                                       & \multicolumn{1}{c|}{P, J}            & \multicolumn{1}{c|}{\begin{tabular}[c]{@{}c@{}}R, Py, Js, C\#,\\ Go, C, C++\end{tabular}}          \\ \hline
\textbf{Animation}                     & \multicolumn{1}{l|}{}                                                                              & \begin{tabular}[c]{@{}c@{}}R, T, Py, P, Js, J,\\ C\#, Go, C, C++\end{tabular} & \multicolumn{1}{c|}{\begin{tabular}[c]{@{}c@{}}R, T, P, Js,\\ J, C\#, Go, C, C++\end{tabular}}     & \multicolumn{1}{c|}{Py}                                                                      & \multicolumn{1}{l|}{}               &                                                                                                    \\ \hline
\textbf{Video}                         & \multicolumn{1}{l|}{}                                                                              & \begin{tabular}[c]{@{}c@{}}R, T, Py, P, Js, J,\\ C\#, Go, C, C++\end{tabular} & \multicolumn{1}{c|}{\begin{tabular}[c]{@{}c@{}}R, P, J, C\#,\\ Go, C, C++\end{tabular}}            & \multicolumn{1}{c|}{T,Py,Js}                                                                 & \multicolumn{1}{l|}{}               &                                                                                                    \\ \hline
\textbf{Table}                         & \multicolumn{1}{l|}{}                                                                              & \begin{tabular}[c]{@{}c@{}}R, T, Py, P, Js, J,\\ C\#, Go, C, C++\end{tabular} & \multicolumn{1}{c|}{\begin{tabular}[c]{@{}c@{}}R, T, Py, P, Js, J,\\ C\#, Go, C, C++\end{tabular}} & \multicolumn{1}{l|}{}                                                                        & \multicolumn{1}{l|}{}               &                                                                                                    \\ \hline
\textbf{Github}                        & \multicolumn{1}{c|}{\begin{tabular}[c]{@{}c@{}}R, T, Py, P, Js, J,\\ C\#, Go, C, C++\end{tabular}} & \multicolumn{1}{l|}{}                                                         & \multicolumn{1}{l|}{}                                                                              & \multicolumn{1}{c|}{}                                                                        & \multicolumn{1}{c|}{T, P}           & \multicolumn{1}{c|}{\begin{tabular}[c]{@{}c@{}}R, Py, Js, J, \\ C\#, Go, C, C++\end{tabular}}      \\ \hline
\textbf{Links}                         & \multicolumn{1}{c|}{\begin{tabular}[c]{@{}c@{}}R, T, Py, P, Js, J,\\ C\#, Go, C, C++\end{tabular}} & \multicolumn{1}{l|}{}                                                         & \multicolumn{1}{l|}{}                                                                              & \multicolumn{1}{l|}{}                                                                        & \multicolumn{1}{l|}{}               & \multicolumn{1}{c|}{\begin{tabular}[c]{@{}c@{}}R, T, Py, P, Js,\\ J, C\#, Go, C, C++\end{tabular}} \\ \hline
\textbf{Project}                       & \multicolumn{1}{c|}{Js, Go, C++}                                                                   & \begin{tabular}[c]{@{}c@{}}R, T, Py, P, \\ J, C\#, C\end{tabular}             & \multicolumn{1}{c|}{R, T, P, C}                                                                    & \multicolumn{1}{c|}{Py, J, Go, C++}                                                          & \multicolumn{1}{c|}{Js, C\#}        &                                                                                                    \\ \hline
\textbf{Inline}                        & \multicolumn{1}{c|}{\begin{tabular}[c]{@{}c@{}}R, T, Py, P, Js, J,\\ C\#, Go, C, C++\end{tabular}} & \multicolumn{1}{l|}{}                                                         & \multicolumn{1}{c|}{P}                                                                             & \multicolumn{1}{c|}{\begin{tabular}[c]{@{}c@{}}R, T, Py, Js,\\ C\#, Go, C, C++\end{tabular}} & \multicolumn{1}{c|}{}               &                                                                                                    \\ \hline
\textbf{Code}                          & \multicolumn{1}{c|}{T, Py}                                                                         & \begin{tabular}[c]{@{}c@{}}R, P, Js, J,\\ C\#, Go, C, C++\end{tabular}        & \multicolumn{1}{c|}{\begin{tabular}[c]{@{}c@{}}R, P, J,\\ C\#, Go, C\end{tabular}}                 & \multicolumn{1}{c|}{Py, Js, C++}                                                             & \multicolumn{1}{c|}{T}              &                                                                                                    \\ \hline
\end{tabular}
\caption{Wilcoxon p-value and Cliff's delta for features being considered, across each of the ten programming languages considered - C, C++, C\#,R: Ruby, T: Typescript, Py: Python, P: PHP, Js: Javascript, J: Java}
\label{tab:wilcoxon_cliff}
\end{table*}

We start the statistical analysis with a null hypothesis that both the groups of data follow same statistical distribution. We then assign ranks to each repository in the groups, based on the feature value, considering one feature at once. The means of ranks are calculated for each group and the difference in the mean values is processed to obtain the p-value. The z-score, or the standard deviation of the groups is calculated, assuming a near-normal distribution among the samples. The probability (p-value) is then determined based on the z-score.
Considering the confidence level to be 95\%, we compare the p-value with $\alpha$ (0.05). According to Wilcoxon's rank sum test, if the p-value is less than $\alpha$, then the null hypothesis is rejected, indicating that both the groups differ in their distribution. The wilcoxon's rank sum test is thus applied for all the ten features under consideration, for each of the programming languages. While the p-value ranges have differed across different programming languages, the results of Wilcoxon's rank sum test indicate that three of the ten features considered had p-value greater than 0.05 for all programming languages and three other features had p-value less than 0.05 for all the programming languages.



%

To further understand the extent of impact of each feature on the popularity of project, beyond the p-value, we apply Cliff's delta on both the data groups, as it is observed to well support the hypothesis testing method \cite{macbeth2011cliff}. As a part of applying Cliff's delta on the popular and non-popular repository groups, the probabilities of a feature value of a random repository in a specific group being greater than the feature value of a repository in another group, and vice versa are calculated and their difference is obtained. The difference in probabilities is termed as cliff's delta and the value determines if the association between the feature and popularity is small, medium, large or negligible. 


The result of applying wilcoxon's rank sum test and cliff's delta on the popular and non-popular groups of data, for each feature is presented in Table \ref{tab:wilcoxon_cliff}.



\begin{table}[]
\begin{tabular}{|l|rl|}
\hline
             & \multicolumn{2}{r|}{\textbf{Fisher's Exact test p-value}}                                                                                                                                                 \\ \hline
\textbf{Category}     & \multicolumn{1}{r|}{\textbf{\textless{}0.05}}                                                       & \textbf{\textgreater{}0.05}                                                                         \\ \hline
\textbf{What}         & \multicolumn{1}{r|}{\begin{tabular}[c]{@{}r@{}}C, C++, C\#, Go,\\ J,  Js, P, Py, R, T\end{tabular}} &                                                                                                     \\ \hline
\textbf{Why}          & \multicolumn{1}{l|}{}                                                                               & \multicolumn{1}{r|}{\begin{tabular}[c]{@{}r@{}}C, C++, C\#, Go,\\ J,  Js, P, Py, R, T\end{tabular}} \\ \hline
\textbf{How}          & \multicolumn{1}{l|}{C\#}                                                                            & \begin{tabular}[c]{@{}l@{}}C, C++, Go, J, \\ Js, P, Py, R, T\end{tabular}                           \\ \hline
\textbf{When}         & \multicolumn{1}{l|}{}                                                                               & \multicolumn{1}{r|}{\begin{tabular}[c]{@{}r@{}}C, C++, C\#, Go,\\ J,  Js, P, Py, R, T\end{tabular}} \\ \hline
\textbf{Who}          & \multicolumn{1}{r|}{\begin{tabular}[c]{@{}r@{}}C, C++, C\#, Go, \\ Js, P, Py, R, T\end{tabular}}    & J                                                                                                   \\ \hline
\textbf{Reference}    & \multicolumn{1}{r|}{\begin{tabular}[c]{@{}r@{}}C, C++, C\#, Go,\\ J,  Js, P, Py, R, T\end{tabular}} &                                                                                                     \\ \hline
\textbf{Contribution} & \multicolumn{1}{r|}{\begin{tabular}[c]{@{}r@{}}C, C++, C\#, Go,\\ J,  Js, P, Py, R, T\end{tabular}} &                                                                                                     \\ \hline
\end{tabular}
\caption{Fishers exact test p-value for categories being considered}
\label{tab:fishers}
\vspace{-10mm}
\end{table}

\subsubsection{Category Analysis}
The categories present in readme file of each repository have been identified. The presence or absence of a category is noted by a boolean value, where `0' indicates absence of the specific category and `1' indicates the presence of the category. We apply statistical hypothesis tests to understand the impact of each category on the popularity of the repositories in each programming language. While Wilcoxon's rank sum test could identify the impact in case of numerical data, the current analysis supports categorical and non-numerical data. Hence, we apply Fisher's exact test on the repositories for each of the categories to understand if there exists a significant association between presence of a category in the readme file and the popularity group of the repository. The Fisher's exact test forms a 2X2 contingency matrix, with the count of popular and non-popular repositories comprising of a category, and not comprising of the category respectively. This contingency table is used to calculate the hyper-geometric probability value  (p-value) of the category over two groups of data. 

Considering the confidence level to be 95\%, the categories with p-value less than 0.05 are considered to be significantly associated with  the popularity of the repositories. The p-value for three and two categories were observed to be less than 0.05 and greater than 0.05 respectively for all programming languages. The results of the Fisher's exact test are presented in Table \ref{tab:fishers}. 



\subsubsection{Identifying Important Features and Categories}
To understand the important features and categories that could be considered as strongly associated with popularity of a repository in a specific programming language, a random forest classifier is built individually for set of features and categories. The features and categories that resulted in better accuracy of the random forest classifier were then considered as important features and categories. 

To avoid reduced importance of correlated features in the random forest classifier, we first check for the presence of collinearity among the features using Spearman's correlation factor. For every pair of features, the spearman's correlation factor $\rho$ is calculated and compared against the threshold. The threshold for $\rho$ has been set to 0.7, inline with the threshold in the study by Fan et al \cite{fan2021makes}.

While the spearman's correlation factor varied for pairs of features across different programming languages, the correlation factor for \textit{github links} and \textit{external links}, was observed to be greater than the threshold for all programming languages. The list of correlated features across the ten programming languages considered is presented in Table \ref{tab:corrfeatures}. If there is only one pair of correlating features, we randomly drop one of those features in building the random forest classifier. If there are more than one pair of correlating features, we identify if there exists a common feature among these pairs, retain this feature, and discard the rest of the features. In cases where there does not exist any common feature among the pairs, one of the feature in the pairs is randomly discarded, and the random forest classifier is built using the rest of the features.

\begin{table}[]
\begin{tabular}{|l|l|}
\hline
\textbf{Language}   & \textbf{Correlating features}                              \\ \hline
C          & {[}\{Github, Links\}{]}                                    \\ \hline
C++        & {[}\{Github, Links\}, \{Image, Links\}{]}                  \\ \hline
C\#        & {[}\{Github, Links\}, \{Image, Links\}{]}                  \\ \hline
Go         & {[}\{Github, Links\}, \{Image, Links\}, \{List, Links\}{]} \\ \hline
Java       & {[}\{Github, Links\}{]}                                    \\ \hline
Javascript & {[}\{Github, Links\}, \{Image, Links\}{]}                  \\ \hline
PHP        & {[}\{Github, Links\}{]}                                    \\ \hline
Python     & {[}\{Github, Links\}{]}                                    \\ \hline
Ruby       & {[}\{Github, Links\}{]}                                    \\ \hline
Typescript          & {[}\{Github, Links\}{]}                                    \\ \hline
\end{tabular}
\caption{Correlating features across the ten programming languages}
\label{tab:corrfeatures}
\vspace{-10mm}
\end{table}
The spearman's correlation factor was also calculated for all the pairs of categories. However, no two categories were observed to have strong correlation, with $\rho$ value less than 0.7 for all the pairs across all the programming languages. 


The random forest classifier is built using the non-correlating features, and a train to test ration of 70:30. While the train to test ratio remained constant, the test and train data was altered with replacement using the out-of-bag bootstrap strategy, for 1000 iterations. As a result, the accuracy of random forest classifier in each iteration has been calculated and the average accuracy of the model was obtained. We observed the average accuracy of random forest classifier built to be at least 70\% for nine of the ten programming languages, while that of PHP was observed to be 63\%. We calculate the importance of each feature using the gini importance \cite{breiman1996some, breiman2001random} that could be obtained from the random forest classifier model. The gini importance ranks the features in terms of the extent to which a feature decreases impurity of the model. 

The average of decrease in impurity corresponding to a feature provides insights on the importance of the feature. Any possible incorrect interpretations are mitigated by considering the variance in accuracy of the model with random shuffling of each feature or category using permutation-based importance \cite{cutler2012random}. 
A larger variance in accuracy of the model implies more importance of the feature or category with respect to popularity.  
Both methods to assess importance of a feature revealed similar levels of importance for majority of the features and categories. If there are any discrepancies in levels of importance across two methods, the difference in impurity values of the respective features or categories and the difference in accuracy variance of the respective features or categories is calculated and compared. If there is a larger difference in the accuracy variance values, then the respective features are ranked based on permutation-based importance, else, they are ranked based on gini importance. The ranks assigned to the features and categories in each of the programming languages are presented in Tables \ref{tab:feature-rank} and \ref{tab:catrank}.

For example, a discrepancy in the levels of importance was observed in case of the \textit{Go} programming language where \textit{Inline} and \textit{Code} features were ranked as the fourth and fifth most important by gini importance, and vice-versa by permutation-based importance. However, the difference in the importance scores were considerably low and almost zero in case of permutation-based importance, when rounded up to three decimals. Hence, \textit{Inline} feature is ranked higher than the \textit{Code} feature.

The correlating features identified through spearman's correlation analysis were assigned the same rank as that of the corresponding correlating feature. 
For example, since \textit{Links} and \textit{GitHub}, and the \textit{Links} and \textit{Images} features were observed to have stronger correlation in C++, the random forest classifier built for C++ considers only \textit{Links} feature and thus the results of gini-importance and permutation-based importance do not result in the values for \textit{Images} and \textit{Github}. However, since these two features are strongly correlated to \textit{Links} feature, the values of importance and the ranks could be same as the \textit{Links} feature.
The results of ranking, in the descending order of importance of features is presented in Table \ref{tab:feature-rank}.

A similar approach, as for features, was followed to build the random forest classifier based on categories. The train to test ratio was considered to be 70:30, with out-of-bag bootstrap strategy applied over 1000 iterations. The average accuracy of the random forest classifier built based on categories was observed to be at least 72\% for all the programming languages, except for PHP and Ruby, for which, the accuracy was 64\%. The gini-importance and permutation-based importance was calculated for each of the categories, and the level of importance of categories was observed to be the same across both the methods. The result of ranking the categories based on their importance is presented in Table \ref{tab:catrank}.

    
\begin{table}[]
\begin{tabular}{|c|cccccccccc|}
\hline
                & \multicolumn{10}{c|}{\textbf{Features with their corresponding ranks of importance}}                                                                                                                                                                                                                                                                                                                                             \\ \hline
                & \multicolumn{1}{c|}{\textbf{GitHub}} & \multicolumn{1}{c|}{\textbf{Links}} & \multicolumn{1}{c|}{\textbf{Image}} & \multicolumn{1}{c|}{\textbf{List}} & \multicolumn{1}{c|}{\textbf{Inline}} & \multicolumn{1}{c|}{\textbf{Code}} & \multicolumn{1}{c|}{\textbf{Video}} & \multicolumn{1}{c|}{\textbf{Project}} & \multicolumn{1}{c|}{\textbf{Table}} & \textbf{Animation} \\ \hline
\textbf{C}      & \multicolumn{1}{c|}{1}               & \multicolumn{1}{c|}{1}              & \multicolumn{1}{c|}{2}              & \multicolumn{1}{c|}{3}             & \multicolumn{1}{c|}{4}               & \multicolumn{1}{c|}{5}             & \multicolumn{1}{c|}{6}              & \multicolumn{1}{c|}{7}                & \multicolumn{1}{c|}{8}              & 9                  \\ \hline
\textbf{C++}    & \multicolumn{1}{c|}{1}               & \multicolumn{1}{c|}{1}              & \multicolumn{1}{c|}{2}              & \multicolumn{1}{c|}{3}             & \multicolumn{1}{c|}{4}               & \multicolumn{1}{c|}{5}             & \multicolumn{1}{c|}{7}              & \multicolumn{1}{c|}{6}                & \multicolumn{1}{c|}{9}              & 8                  \\ \hline
\textbf{C\#}    & \multicolumn{1}{c|}{1}               & \multicolumn{1}{c|}{1}              & \multicolumn{1}{c|}{1}              & \multicolumn{1}{c|}{3}             & \multicolumn{1}{c|}{5}               & \multicolumn{1}{c|}{7}             & \multicolumn{1}{c|}{8}              & \multicolumn{1}{c|}{4}                & \multicolumn{1}{c|}{9}              & 6                  \\ \hline
\textbf{Go}     & \multicolumn{1}{c|}{1}               & \multicolumn{1}{c|}{1}              & \multicolumn{1}{c|}{1}              & \multicolumn{1}{c|}{1}             & \multicolumn{1}{c|}{4}               & \multicolumn{1}{c|}{5}             & \multicolumn{1}{c|}{7}              & \multicolumn{1}{c|}{6}                & \multicolumn{1}{c|}{8}              & 8                  \\ \hline
\textbf{Java}   & \multicolumn{1}{c|}{1}               & \multicolumn{1}{c|}{1}              & \multicolumn{1}{c|}{3}              & \multicolumn{1}{c|}{2}             & \multicolumn{1}{c|}{4}               & \multicolumn{1}{c|}{6}             & \multicolumn{1}{c|}{8}              & \multicolumn{1}{c|}{5}                & \multicolumn{1}{c|}{8}              & 7                  \\ \hline
\textbf{Javascript}     & \multicolumn{1}{c|}{1}               & \multicolumn{1}{c|}{1}              & \multicolumn{1}{c|}{1}              & \multicolumn{1}{c|}{3}             & \multicolumn{1}{c|}{6}               & \multicolumn{1}{c|}{4}             & \multicolumn{1}{c|}{6}              & \multicolumn{1}{c|}{5}                & \multicolumn{1}{c|}{6}              & 6                  \\ \hline
\textbf{PHP}    & \multicolumn{1}{c|}{1}               & \multicolumn{1}{c|}{1}              & \multicolumn{1}{c|}{2}              & \multicolumn{1}{c|}{3}             & \multicolumn{1}{c|}{4}               & \multicolumn{1}{c|}{5}             & \multicolumn{1}{c|}{7}              & \multicolumn{1}{c|}{6}                & \multicolumn{1}{c|}{8}              & 8                  \\ \hline
\textbf{Python} & \multicolumn{1}{c|}{1}               & \multicolumn{1}{c|}{1}              & \multicolumn{1}{c|}{4}              & \multicolumn{1}{c|}{2}             & \multicolumn{1}{c|}{3}               & \multicolumn{1}{c|}{5}             & \multicolumn{1}{c|}{8}              & \multicolumn{1}{c|}{6}                & \multicolumn{1}{c|}{8}              & 7                  \\ \hline
\textbf{Ruby}   & \multicolumn{1}{c|}{1}               & \multicolumn{1}{c|}{1}              & \multicolumn{1}{c|}{2}              & \multicolumn{1}{c|}{3}             & \multicolumn{1}{c|}{4}               & \multicolumn{1}{c|}{5}             & \multicolumn{1}{c|}{7}              & \multicolumn{1}{c|}{6}                & \multicolumn{1}{c|}{7}              & 7                  \\ \hline
\textbf{Typescript}      & \multicolumn{1}{c|}{1}               & \multicolumn{1}{c|}{1}              & \multicolumn{1}{c|}{4}              & \multicolumn{1}{c|}{6}             & \multicolumn{1}{c|}{3}               & \multicolumn{1}{c|}{2}             & \multicolumn{1}{c|}{5}              & \multicolumn{1}{c|}{7}                & \multicolumn{1}{c|}{9}              & 7                  \\ \hline
\end{tabular}
\caption{Feature Importance and ranking based on gini-importance and permutation-based importance across the 10 programming languages. Ranks of importance are presented in descending order, where feature with rank 1 has highest importance and that with rank 9 has the least importance.}
\label{tab:feature-rank}
\end{table}

\begin{table}[]
\begin{tabular}{|c|ccccccc|}
\hline
                & \multicolumn{7}{c|}{\textbf{Categories with their corresponding ranks of importance}}                                                                                                                                                                                                              \\ \hline
                & \multicolumn{1}{c|}{\textbf{What}} & \multicolumn{1}{c|}{\textbf{Why}} & \multicolumn{1}{c|}{\textbf{How}} & \multicolumn{1}{c|}{\textbf{When}} & \multicolumn{1}{c|}{\textbf{Who}} & \multicolumn{1}{c|}{\textbf{Refernece}} & \textbf{Contribution} \\ \hline
\textbf{C}      & \multicolumn{1}{c|}{4}             & \multicolumn{1}{c|}{7}            & \multicolumn{1}{c|}{5}            & \multicolumn{1}{c|}{6}             & \multicolumn{1}{c|}{3}            & \multicolumn{1}{c|}{1}                  & 2                     \\ \hline
\textbf{C++}    & \multicolumn{1}{c|}{4}             & \multicolumn{1}{c|}{7}            & \multicolumn{1}{c|}{6}            & \multicolumn{1}{c|}{5}             & \multicolumn{1}{c|}{2}            & \multicolumn{1}{c|}{3}                  & 1                     \\ \hline
\textbf{C\#}    & \multicolumn{1}{c|}{4}             & \multicolumn{1}{c|}{7}            & \multicolumn{1}{c|}{5}            & \multicolumn{1}{c|}{6}             & \multicolumn{1}{c|}{3}            & \multicolumn{1}{c|}{2}                  & 1                     \\ \hline
\textbf{Go}     & \multicolumn{1}{c|}{2}             & \multicolumn{1}{c|}{7}            & \multicolumn{1}{c|}{6}            & \multicolumn{1}{c|}{5}             & \multicolumn{1}{c|}{4}            & \multicolumn{1}{c|}{3}                  & 1                     \\ \hline
\textbf{Java}   & \multicolumn{1}{c|}{3}             & \multicolumn{1}{c|}{7}            & \multicolumn{1}{c|}{5}            & \multicolumn{1}{c|}{4}             & \multicolumn{1}{c|}{6}            & \multicolumn{1}{c|}{1}                  & 2                     \\ \hline
\textbf{Javascript}     & \multicolumn{1}{c|}{4}             & \multicolumn{1}{c|}{7}            & \multicolumn{1}{c|}{6}            & \multicolumn{1}{c|}{5}             & \multicolumn{1}{c|}{2}            & \multicolumn{1}{c|}{3}                  & 1                     \\ \hline
\textbf{PHP}    & \multicolumn{1}{c|}{4}             & \multicolumn{1}{c|}{7}            & \multicolumn{1}{c|}{5}            & \multicolumn{1}{c|}{5}             & \multicolumn{1}{c|}{1}            & \multicolumn{1}{c|}{2}                  & 3                     \\ \hline
\textbf{Python} & \multicolumn{1}{c|}{4}             & \multicolumn{1}{c|}{7}            & \multicolumn{1}{c|}{5}            & \multicolumn{1}{c|}{5}             & \multicolumn{1}{c|}{3}            & \multicolumn{1}{c|}{1}                  & 2                     \\ \hline
\textbf{Ruby}   & \multicolumn{1}{c|}{5}             & \multicolumn{1}{c|}{7}            & \multicolumn{1}{c|}{4}            & \multicolumn{1}{c|}{5}             & \multicolumn{1}{c|}{3}            & \multicolumn{1}{c|}{1}                  & 2                     \\ \hline
\textbf{Typescript}      & \multicolumn{1}{c|}{4}             & \multicolumn{1}{c|}{7}            & \multicolumn{1}{c|}{5}            & \multicolumn{1}{c|}{5}             & \multicolumn{1}{c|}{1}            & \multicolumn{1}{c|}{3}                  & 2                     \\ \hline
\end{tabular}
\caption{Category Importance and ranking based on gini-importance and permutation-based importance across the 10 programming languages. Ranks of importance are presented in descending order, where category with rank 1 has highest importance and that with rank 7 has the least importance. }
\label{tab:catrank}
\end{table}
\section{Results}
\label{res}
In this section we elaborate the results of the analysis and answer the research questions based on the results.

\textbf{RQ1: Does presence of a specific feature relate to the popularity of the repository?}
\begin{figure}
    \centering
    \includegraphics[scale = 0.7]{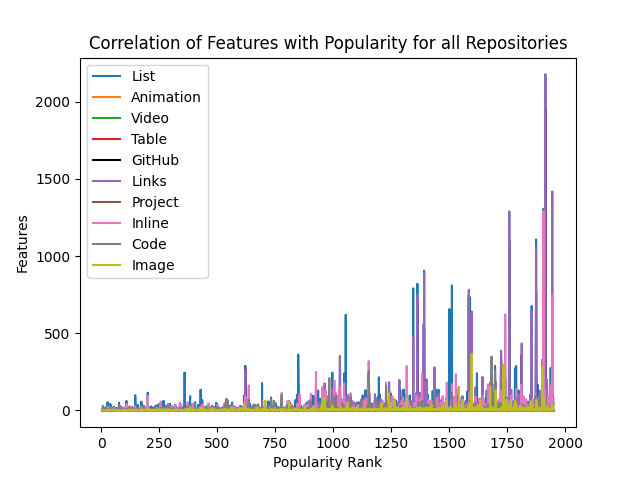}
    \caption{Plot displaying correlation of all 10 features across repositories in all the 10 programming languages}
    \label{fig:all_features}
\end{figure}

\begin{figure}
\centering
\begin{minipage}{.5\linewidth}
  \centering
  \includegraphics[width=.9\linewidth]{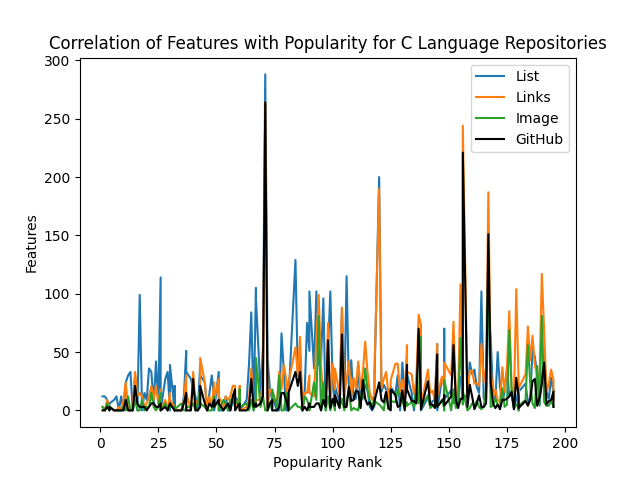}
  \captionof{figure}{Plot displaying correlation of four features\\ with popularity for C language repositories}
  \label{fig:features_C}
\end{minipage}%
\begin{minipage}{.5\linewidth}
  \centering
  \includegraphics[width=.9\linewidth]{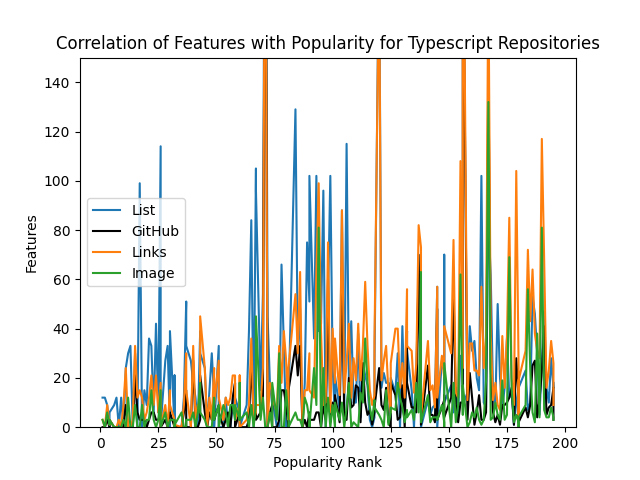}
  \captionof{figure}{Plot displaying correlation of four features with popularity of Typescript language repositories}
  \label{fig:features_ts}
\end{minipage}
\end{figure}

The features listed in Table \ref{tab:features} have been extracted from each of the readme files of the 2000 repositories. The count of these features is considered to be associated with project popularity. {The distribution of feature counts for all the repositories under consideration, irrespective of their programming languages, with respect to the popularity rank is displayed in Fig. \ref{fig:all_features}. The popularity rank, here is calculated across all the 200 repositories. The spike in the graph in Fig. \ref{fig:all_features} indicates the count of the corresponding feature. Higher and larger number of spikes for features with increasing popularity rank indicate that the respective feature could have a positive correlation with the popularity rank, i.e., the count of the feature increases with increase in popularity rank. 
The distribution of \textit{List}, \textit{GitHub}, \textit{Links} and \textit{Image} feature counts with respect to popularity rank of the repositories in C and Typescript languages is presented in Fig. \ref{fig:features_C} and Fig. \ref{fig:features_ts} respectively. Fig. \ref{fig:features_C} depicts increasing number of spikes for \textit{GitHub} and \textit{Links} features, with popularity rank. The spikes for \textit{Image} and \textit{List} features are increasing at a slower rate than \textit{GitHub} and \textit{Links}.} 

{Also, the \textit{List} feature has spikes increasing unevenly with popularity rank. This could indicate that \textit{GitHub} and \textit{Links} features have more positive correlation with popularity rank, followed by \textit{Image} and \textit{List} features for repositories in C language. Fig. \ref{fig:features_ts} depicts increasing number of spikes for \textit{GitHub} and \textit{Links} features with popularity rank, while that of \textit{Image} and \textit{List} features do not evenly increase unlike \textit{Links} and \textit{GitHub} features, with more unevenness in case of \textit{List} feature. This could indicate strong positive correlation of \textit{GitHub} and \textit{Links} features with popularity ranks, and weaker correlation of \textit{Image}, followed by \textit{List} features. Also, the Fig. \ref{fig:features_C} and Fig. \ref{fig:features_ts} indicate that \textit{List} feature in C language repositories has more positive correlation than the \textit{List} feature in Typescript language repositories, as there are more number of even spikes for \textit{List} feature in Fig. \ref{fig:features_C}, than in Fig. \ref{fig:features_ts}. The graphs presented here are intended to bring-out a high level idea of correlation of some features with popularity rank. Exact values of all feature counts for repositories in all the languages can be found here\footnote{\url{https://osf.io/feutq/?view_only=e5517387c69f4d959a852758a085ea25}}.}
However, to gain a much clearer understanding statistically, if each of the features have a correlation with the popularity of the repositories, we applied non-parametric statistical hypothesis tests that determine the statistical significance between two groups of data. Such tests were observed to be applied in literature in studies with similar aims of understanding impact of specific nature of projects on the projects' characteristics and the association between these characteristics and project nature \cite{fan2021makes, chowdhury2016characterizing, fan2018chaff, fan2019impact, fan2018early}.

Wilcoxon's rank-sum test was applied on each of the ten features to identify if there exists a statistically significant difference in the feature values among the popular and non-popular groups of repositories. The p-value of wilcoxon's rank sum test determines the presence of significant statistical difference.  Features with p-value less than 0.05 ($\alpha$) imply that the respective features are associated with popularity of the repository and are observed to vary across popular and non-popular repositories. 

It has been observed that \textit{Image}, \textit{Github}, \textit{Links} and \textit{Inline} features are strongly associated with popularity of repositories and \textit{Animated Image}, \textit{Video} and \textit{Table} were observed not to have a strong relation with popularity of repositories across all programming languages, based on the corresponding p-value ranges, as presented in Table \ref{tab:wilcoxon_cliff}. This indicates that the features with strong association could be considered as differentiating factors, with respect to readme files, among the popular and non-popular repositories. \textit{List} feature was observed to be strongly associated with popularity across the 10 programming languages, except for PHP and Typescript. Also, the \textit{Code} feature was observed not to have a strong relation with popularity across the 10 programming languages, except for Python and Typescript.


The results were further supported by Cliff's delta test, which also revealed the extent of effect of a feature on the popularity of the repository. The \textit{Links} feature was observed to have larger effect size, indicating strongest association with popularity across all programming languages. Also, the \textit{Image} and \textit{Github} features had larger effect sizes for all the programming languages, except for Java, PHP, Typescript and PHP, Typescript respectively. It was also observed that the \textit{Animated Image}, \textit{Video}, and \textit{Table} features had almost negligible association with the popularity of the repositories, irrespective of the programming language corresponding to the repository. We observed that majority of the repositories, in both popular and non-popular categories did not contain \textit{Animated Images} or \textit{Tables}. 

For example, only 9 of the 200 repositories corresponding to C language contained at least one \textit{Animated Image}, with two being the maximum number of \textit{Animated Images} present in any repository. Two C\# repositories with popularity ranks 124 and 39, belonging to popular and non-popular groups respectively contained four \textit{Animated Images} each, in their readme files, thus, indicating almost no relation of \textit{Animated Images} with the popularity. Also, only one repository with popularity rank 190 among the 200 Typescript repositories contained tables in the readme files, thus making it difficult to identify distinguishing patterns among popular and non-popular repositories with respect to \textit{Tables}.
\newline 
\newline
\fbox{\begin{minipage}{46em}
Inclusion of links to external sources to provide easy navigation to the references in readme files is observed as a common practice in popular repositories across all programming languages. Also, including \textit{Code} in the readme files of was observed to be prevalent in the most popular Typescript projects.


\end{minipage}}
\newline

\textbf{RQ2: Does presence of a specific category relate to the popularity of the repository?}
\begin{figure}
\centering
\begin{minipage}{.5\linewidth}
  \centering
  \includegraphics[width=.9\linewidth]{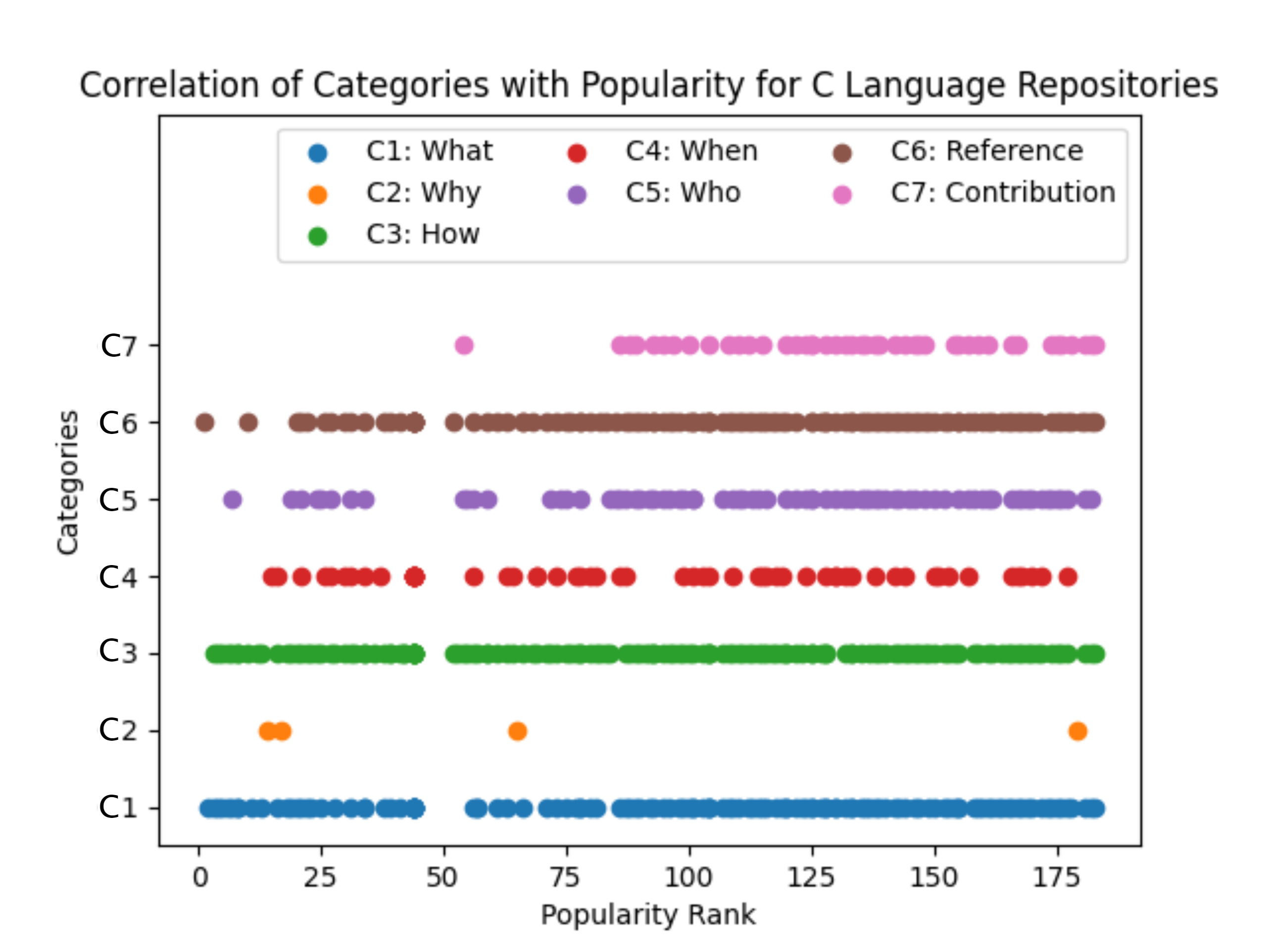}
  \captionof{figure}{Plot displaying correlation of the seven categories\\ with popularity for C language repositories}
  \label{fig:Categories_C}
\end{minipage}%
\begin{minipage}{.5\linewidth}
  \centering
  \includegraphics[width=.9\linewidth]{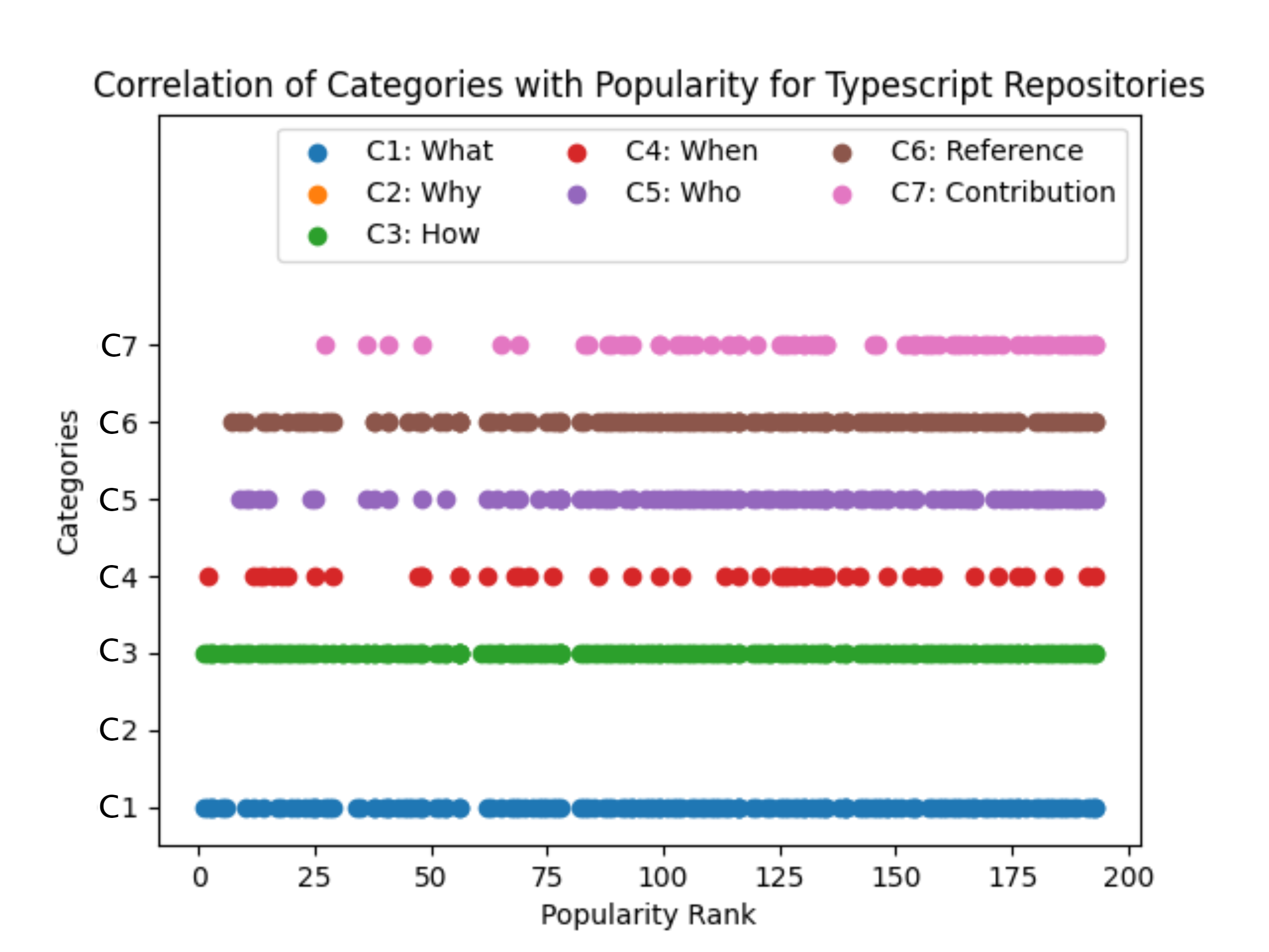}
  \captionof{figure}{Plot displaying correlation of the seven categories with popularity of Typescript language repositories}
  \label{fig:Categories_ts}
\end{minipage}
\end{figure}

The content in the readme files has been processed to identify the presence of categories listed in Table \ref{tab:categories}. Every repository has been appended with boolean values corresponding to each of the categories, where `0' and `1' indicate the absence and presence of the categories in the readme file. {The distribution of presence of the seven categories across repositories in C and Typescript languages are presented in Fig. \ref{fig:Categories_C} and Fig. \ref{fig:Categories_ts} respectively. The increasing density of dots in the plots indicates increasing presence of the categories. The presence of \textit{Reference}, \textit{Contribution} and \textit{Who} categories is observed to be increasing with popularity rank, compared to other features, in Fig. \ref{fig:Categories_C}, which could indicate strong positive correlation of these features with popularity rank for C language repositories. Moreover, \textit{Reference} category has more density than \textit{Contribution}, which has more density than \textit{Who} category, towards increasing popularity rank, which could indicate that correlation of these categories with popularity rank is in the order of \textit{Reference} > \textit{Contribution} > \textit{Who} for C language repositories. Similarly, in case of Typescript repositories, the densities of \textit{Reference}, \textit{Contribution} and \textit{Who} categories are increasing with popularity rank, but the \textit{Reference} category also has higher densities at few instances, for lower popularity ranks. The \textit{Who} category has lower density for low popularity rank, and almost continuous density as the popularity rank increases, when compared to \textit{Reference} category. This could indicate that \textit{Who} category has stronger correlation with popularity rank, than that of \textit{Reference} category for repositories in Typescript language. The presence of categories for repositories in each programming language can be found here\footnote{\url{https://osf.io/feutq/?view_only=e5517387c69f4d959a852758a085ea25}}.
}

To gain a clearer understanding of the correlation, similar to the case of features, we employ the statistical analysis method, Fisher's exact test. Fisher's exact test supports analysis of \textit{null hypothesis independence} in samples with non-numeric categorical values, and hence, was applied to assess statistical significance of a category in contributing to the popularity. 
The p-value less than 0.05 ($\alpha$) implies that the category and the popularity of the repository are not independent of each other and thus, the category has a strong association with the popularity of the repository. We observe that p-values of three of the seven categories are less than 0.05, for all the 10 programming languages, while that of the \textit{Why} and \textit{When} categories is greater than 0.05, for all the 10 programming languages considered, as presented in Table \ref{tab:fishers}. The \textit{Who} category had p-value less than 0.05 for all programming languages except Java and the \textit{How} category had p-value greater than 0.05 for all programming languages except C\#. 

The \textit{How} category was observed to be present in majority of the repositories, across all programming languages considered, except for C\#, irrespective of their popularity. 189 out of the 200 Java repositories contained \textit{How} category in the readme files. This could be the reason for weaker association of \textit{How} with the popularity of the repositories. However, we cannot conclude that the absence of \textit{How} category could change the popularity of the repository.
\newline
\newline
\fbox{\begin{minipage}{46em}
Details corresponding to project functionalities, guidelines to contribute to project and reference sources in readme files are positively and strongly associated with the popularity of repositories in majority of the programming languages. The details of installation instructions or ways to use the repository are present more commonly in popular C\# repositories, when compared to other programming languages.

\end{minipage}}
\newline

\textbf{RQ3: Which features and categories vary among the popular and non-popular repositories?}

The features and categories that have stronger association with popularity among others were identified through analysis of increased impurity using gini importance. The higher the impurity in absence of a feature or category, stronger the association of the respective feature or category on the popularity of the repository. 

Based on the ranks assigned considering results of gini and permutation-based importance, as presented in Table \ref{tab:feature-rank}, we observe that the  \textit{Links} and \textit{Github} features and the \textit{Contribution} and \textit{Reference} features have the strongest association with popularity of the repositories in almost all the 10 programming languages considered. 
The \textit{Image}, \textit{List}, \textit{Inline}, \textit{Code} features follow the \textit{Links} in level of importance on the project popularity. 
The categories \textit{Who}, \textit{What} and \textit{When} follow the \textit{Contribution} and \textit{Reference} categories in the order of extent of association with the project popularity, as seen in Table \ref{tab:catrank}. 
\newline
\newline
\fbox{\begin{minipage}{46em}


The projects with better popularity are observed to contain content referring to the project page, contribution guidelines and are well-organised in the form of lists, images and inline code.  Also, the presence of team information was observed to be a distinguishing factor among the popular and non-popular repositories in PHP and Typescript languages.
\end{minipage}}
\newline

\section{Threats to Validity}
\label{ttv}
In this section we specify the threats to validity with respect to the empirical study performed, the data considered and the statistical methods used to perform the empirical study.

\textbf{\textit{Construct Validity -}}
The features and categories considered for analysis could be valid for all projects across GitHub, irrespective of the programming language or type of the project. However, the results corresponding to association of the features or categories with project popularity are restricted to the 2000 repositories considered for the study. 
Replicating the study on a different set of repositories might yield different results. However, as the current set of repositories span across 10 programming languages, with relatively large distribution in terms of the star counts, this set could be considered as a representative sample of all the repositories on GitHub. Hence, a similar association of the features and categories could be observed on other GitHub repositories, suggesting (but not proving) generalizability of the results. 
 
The popularity rank considered for the study is aimed to provide equal weightage to the star\_count, fork\_count, watcher\_count and pull request \_count of the repositories. However, the popularity of the repositories might also depend on other characteristics of repositories such as contributor count, user count and so on. Including other such factors of the repositories and updating the formula to calculate popularity rank accordingly, could alter the repositories in the current popular and non-popular groups. Some of the repositories currently in the popular group might migrate to non-popular group with change in factors considered to calculate the popularity.

\textbf{\textit{Internal Validity -}}
We define features and categories as the structural and type of content present in the readme files respectively. The notion of these definitions is specific to the study and could vary based on the researchers' perceptions. The features and categories considered for the study are based on the existing literature corresponding to readme files. The features proposed by Fan et al. in \cite{fan2021makes} and the categories proposed by Prana et al. in \cite{prana2019categorizing} have been considered. Thus, there is a possibility that there might exist other features or categories in the readme content, which can be explored in future extensions of the study. However, the categories considered were also observed to be a superset of \textit{GitHub recommended guidelines}, indicating low possibility of missing out on the categories.

The association of the features is identified using the p-value obtained by wilcoxon's rank sum test and the delta value obtained using the cliff's delta statistical hypothesis tests. 
While these statistical tests to identify the statistical significance of features were observed to be relevant to the numerical non-parametric data in the current study \cite{fan2018chaff, fan2021makes}, using other statistical hypothesis tests could result in different results. Similarly, while the use of Fisher's exact test on the non-numerical categorical values is relevant to the current data with respect to the categories \cite{fan2021makes}, other statistical analysis methods could result in different results.




\textbf{\textit{External Validity -}}
The results obtained are confined to the version of repositories and their and corresponding software artifacts as on 16 December 2021, as this empirical study is performed on locally downloaded repositories.

The accuracy of categorizing the extracted information into different categories is dependant on the efficiency of the readme classifier provided by Prana et al. in \cite{prana2019categorizing}. Applying a different classifier for the same class labels might yield in different distributions of the classes in the readme files of the repositories. The accuracy of identifying the feature\_count for each of the features is dependant on the automated approach presented in the study, based on the identification marks discussed in \cite{fan2021makes}. Adding different identification marks for the features could result in varied feature\_counts.  For example, for the \textit{Project} feature, the current identification mark is the presence of \textit{project} keyword in the \textit{urls} in html formats of the readme files. The \textit{Project} feature could further be calculated by considering the presence of \textit{project name} in the \textit{urls} in html format. However, such decisions depend on the perception and choices of the researchers. As we are considering the features presented by Fan et al. in \cite{fan2021makes}, we consider the choice of identification marks discussed in \cite{fan2021makes}.


\section{Discussion}
\label{disc}
This empirical study performed on 2000 GitHub repositories across 10 different programming languages provides insights on possible structural and contextual content that could be included in readme files, to ensure better popularity of the repositories. The results of the study lead to a list of good practices in creating readme files, which are observed to be commonly used in popular repositories and are not very common in the non-popular repositories. Following a structured format using lists, including pictorial representations such as images, adding references and providing links to relevant external sources and other GitHub projects, specifying the ways to contribute to projects, mentioning the team details and functional properties of the repository  are identified as some of the good practices in designing readme files. 
However, we can not claim that these factors correspond to an exhaustive set of good practices, and can be further extended based on other factors of the repository such as the domain, type of usage and so on. Also, these factors are specific to the repositories and popularity metrics considered. The popularity metric could be restricted to fewer dimensions such as stars and forks, or could be further enhanced by adding more dimensions such as number of users and contributors. Using a different set of repositories for popular and non-popular groups could yield different results. However, since there is a large difference between the star count of the popular and non-popular repositories, the current set of repositories could be considered as a representative sample, thus indicating generalisability of the results. However, further studies with varied dimensions of popularity and varied set of repositories should be performed to prove the generalisability of results.  


\textbf{\textit{Insights for Practitioners - }}
The \textit{Github} and \textit{Links} features were observed to be the most impactful features across all programming languages. These features were observed to be largely present in popular repositories and were hardly present in non-popular repositories. These two features can be grouped as links to external sources, which could be narrowed down to \textit{References} category. Software practitioners could thus consider including references, with links to external sources and relevant GitHub projects in their readme files. Software practitioners, especially those working with C\#, Go and Javascript projects, could follow the good practices of structuring their readme files using \textit{Lists} and \textit{Images}, along with \textit{References}, towards arriving at better readme file. 
They could further include the team details of the project, specific steps to be followed to contribute to the project in the readme files to support new developers who wish to contribute to projects, thus, facilitating developer onboarding.




\textbf{\textit{Insights for Researchers - }}
The links, images, lists, contribution guidelines, references, team details and functional features of repositories were observed to be strongly, and positively associated with popularity of the repositories. These features could be taken into account in designing approaches that predict popularity of repositories, and also in approaches that improve or generate readme files.
The above mentioned features and categories could be considered as a preliminary set of good practices in designing readme files and thus could be potentially useful in arriving at a taxonomy of good practices in designing the readme files. 
Researchers could further explore the impact of readme content on other characteristics of the projects such as project progress, issue resolution rate, number of contributors and so on, which could further contribute to taxonomy of good practices for readme files.
The impact of content in readme files on the quality of readme files and the quality of documentation on a broader level could also be explored.

    








\section{Conclusion and Future Work}
\label{concl}

We perform a study on 200 GitHub repositories in each of the top-10 most popular programming languages to understand the relation of features and categories with the popularity of the repositories. The features and categories are identified and their association with popularity has been assessed using statistical hypothesis tests. The importance order of each of the features and categories has further been calculated by building a random forest classifier and applying gini importance and permutation-based importance methods on the random forest classifier generated.

We observe that the presence of references, contribution guidelines, information about the project team and the purpose of the project in readme files are strongly and positively associated with popularity of the repositories. Also, structuring the content as lists, adding images, and links to external sources and other GitHub projects was observed to positively and proportionally be related to the popularity of repositories.
Some features and categories such as details about how to use the project were not observed to be related to the popularity, the reason being that they are present in majority of the readme files, irrespective of popularity, and thus such categories or features could be included as frequent practices in creating a readme file. 

We plan to extend this study to identify the impact of these features and categories on other project attributes such as project progress, project quality and so on. We also plan to perform further studies on multiple projects, to identify new features and categories, and extend the list of features and categories being studied. Future research directions for this study could also include analysing correlation of the features and categories across projects in specific domains such as machine learning, deep learning, game engines, games and so on. This analysis could also help in bringing out patterns of features and categories in readme files in projects of specific domains.
This correlation analysis could help in identifying good practices of readme files across multiple domains and various programming languages. 

\bibliographystyle{ACM-Reference-Format}
\bibliography{sample-base}


\begin{thebibliography}{44}


\ifx \showCODEN    \undefined \def \showCODEN     #1{\unskip}     \fi
\ifx \showDOI      \undefined \def \showDOI       #1{#1}\fi
\ifx \showISBNx    \undefined \def \showISBNx     #1{\unskip}     \fi
\ifx \showISBNxiii \undefined \def \showISBNxiii  #1{\unskip}     \fi
\ifx \showISSN     \undefined \def \showISSN      #1{\unskip}     \fi
\ifx \showLCCN     \undefined \def \showLCCN      #1{\unskip}     \fi
\ifx \shownote     \undefined \def \shownote      #1{#1}          \fi
\ifx \showarticletitle \undefined \def \showarticletitle #1{#1}   \fi
\ifx \showURL      \undefined \def \showURL       {\relax}        \fi
\providecommand\bibfield[2]{#2}
\providecommand\bibinfo[2]{#2}
\providecommand\natexlab[1]{#1}
\providecommand\showeprint[2][]{arXiv:#2}

\bibitem[Aggarwal et~al\mbox{.}(2014)]%
        {aggarwal2014co}
\bibfield{author}{\bibinfo{person}{Karan Aggarwal}, \bibinfo{person}{Abram
  Hindle}, {and} \bibinfo{person}{Eleni Stroulia}.}
  \bibinfo{year}{2014}\natexlab{}.
\newblock \showarticletitle{Co-evolution of project documentation and
  popularity within GitHub}. In \bibinfo{booktitle}{\emph{Proceedings of the
  11th Working Conference on Mining Software Repositories}}.
  \bibinfo{pages}{360--363}.
\newblock


\bibitem[Bao et~al\mbox{.}(2019)]%
        {bao2019large}
\bibfield{author}{\bibinfo{person}{Lingfeng Bao}, \bibinfo{person}{Xin Xia},
  \bibinfo{person}{David Lo}, {and} \bibinfo{person}{Gail~C Murphy}.}
  \bibinfo{year}{2019}\natexlab{}.
\newblock \showarticletitle{A large scale study of long-time contributor
  prediction for github projects}.
\newblock \bibinfo{journal}{\emph{IEEE Transactions on Software Engineering}}
  (\bibinfo{year}{2019}).
\newblock


\bibitem[Basili et~al\mbox{.}(1994)]%
        {basili1994goal}
\bibfield{author}{\bibinfo{person}{Victor~R Basili}, \bibinfo{person}{Gianluigi
  Caldiera}, {and} \bibinfo{person}{H~Dieter Rombach}.}
  \bibinfo{year}{1994}\natexlab{}.
\newblock \bibinfo{title}{Goal, question metric paradigm. Encyclopedia of
  Software Engineering, vol. 1}.
\newblock
\newblock


\bibitem[Blei et~al\mbox{.}(2003)]%
        {blei2003latent}
\bibfield{author}{\bibinfo{person}{David~M Blei}, \bibinfo{person}{Andrew~Y
  Ng}, {and} \bibinfo{person}{Michael~I Jordan}.}
  \bibinfo{year}{2003}\natexlab{}.
\newblock \showarticletitle{Latent dirichlet allocation}.
\newblock \bibinfo{journal}{\emph{Journal of machine Learning research}}
  \bibinfo{volume}{3}, \bibinfo{number}{Jan} (\bibinfo{year}{2003}),
  \bibinfo{pages}{993--1022}.
\newblock


\bibitem[Borges et~al\mbox{.}(2016a)]%
        {borges2016predicting}
\bibfield{author}{\bibinfo{person}{Hudson Borges}, \bibinfo{person}{Andre
  Hora}, {and} \bibinfo{person}{Marco~Tulio Valente}.}
  \bibinfo{year}{2016}\natexlab{a}.
\newblock \showarticletitle{Predicting the popularity of GitHub repositories}.
  In \bibinfo{booktitle}{\emph{Proceedings of the The 12th International
  Conference on Predictive Models and Data Analytics in Software Engineering}}.
  \bibinfo{pages}{1--10}.
\newblock


\bibitem[Borges et~al\mbox{.}(2016b)]%
        {borges2016understanding}
\bibfield{author}{\bibinfo{person}{Hudson Borges}, \bibinfo{person}{Andre
  Hora}, {and} \bibinfo{person}{Marco~Tulio Valente}.}
  \bibinfo{year}{2016}\natexlab{b}.
\newblock \showarticletitle{Understanding the factors that impact the
  popularity of GitHub repositories}. In \bibinfo{booktitle}{\emph{2016 IEEE
  International Conference on Software Maintenance and Evolution (ICSME)}}.
  IEEE, \bibinfo{pages}{334--344}.
\newblock


\bibitem[Borges and Valente(2018)]%
        {borges2018s}
\bibfield{author}{\bibinfo{person}{Hudson Borges} {and}
  \bibinfo{person}{Marco~Tulio Valente}.} \bibinfo{year}{2018}\natexlab{}.
\newblock \showarticletitle{What’s in a github star? understanding repository
  starring practices in a social coding platform}.
\newblock \bibinfo{journal}{\emph{Journal of Systems and Software}}
  \bibinfo{volume}{146} (\bibinfo{year}{2018}), \bibinfo{pages}{112--129}.
\newblock


\bibitem[Breiman(1996)]%
        {breiman1996some}
\bibfield{author}{\bibinfo{person}{Leo Breiman}.}
  \bibinfo{year}{1996}\natexlab{}.
\newblock \showarticletitle{Some properties of splitting criteria}.
\newblock \bibinfo{journal}{\emph{Machine learning}} \bibinfo{volume}{24},
  \bibinfo{number}{1} (\bibinfo{year}{1996}), \bibinfo{pages}{41--47}.
\newblock


\bibitem[Breiman(2001)]%
        {breiman2001random}
\bibfield{author}{\bibinfo{person}{Leo Breiman}.}
  \bibinfo{year}{2001}\natexlab{}.
\newblock \showarticletitle{Random forests}.
\newblock \bibinfo{journal}{\emph{Machine learning}} \bibinfo{volume}{45},
  \bibinfo{number}{1} (\bibinfo{year}{2001}), \bibinfo{pages}{5--32}.
\newblock


\bibitem[C{\^{a}}ndido et~al\mbox{.}(2021)]%
        {DBLP:conf/msr/CandidoHAD21}
\bibfield{author}{\bibinfo{person}{Jeanderson C{\^{a}}ndido},
  \bibinfo{person}{Jan Haesen}, \bibinfo{person}{Maur{\'{\i}}cio Aniche}, {and}
  \bibinfo{person}{Arie van Deursen}.} \bibinfo{year}{2021}\natexlab{}.
\newblock \showarticletitle{An Exploratory Study of Log Placement
  Recommendation in an Enterprise System}. In \bibinfo{booktitle}{\emph{18th
  {IEEE/ACM} International Conference on Mining Software Repositories, {MSR}
  2021, Madrid, Spain, May 17-19, 2021}}. \bibinfo{publisher}{{IEEE}},
  \bibinfo{pages}{143--154}.
\newblock
\urldef\tempurl%
\url{https://doi.org/10.1109/MSR52588.2021.00027}
\showDOI{\tempurl}


\bibitem[Chen et~al\mbox{.}(2020)]%
        {chen2020empirical}
\bibfield{author}{\bibinfo{person}{Haowen Chen}, \bibinfo{person}{Xiao-Yuan
  Jing}, \bibinfo{person}{Zhiqiang Li}, \bibinfo{person}{Di Wu},
  \bibinfo{person}{Yi Peng}, {and} \bibinfo{person}{Zhiguo Huang}.}
  \bibinfo{year}{2020}\natexlab{}.
\newblock \showarticletitle{An empirical study on heterogeneous defect
  prediction approaches}.
\newblock \bibinfo{journal}{\emph{IEEE Transactions on Software Engineering}}
  (\bibinfo{year}{2020}).
\newblock


\bibitem[Chowdhury and Hindle(2016)]%
        {chowdhury2016characterizing}
\bibfield{author}{\bibinfo{person}{Shaiful~Alam Chowdhury} {and}
  \bibinfo{person}{Abram Hindle}.} \bibinfo{year}{2016}\natexlab{}.
\newblock \showarticletitle{Characterizing energy-aware software projects: Are
  they different?}. In \bibinfo{booktitle}{\emph{Proceedings of the 13th
  International Conference on Mining Software Repositories}}.
  \bibinfo{pages}{508--511}.
\newblock


\bibitem[Ciniselli et~al\mbox{.}(2021)]%
        {DBLP:conf/msr/CiniselliCPPPB21}
\bibfield{author}{\bibinfo{person}{Matteo Ciniselli}, \bibinfo{person}{Nathan
  Cooper}, \bibinfo{person}{Luca Pascarella}, \bibinfo{person}{Denys
  Poshyvanyk}, \bibinfo{person}{Massimiliano~Di Penta}, {and}
  \bibinfo{person}{Gabriele Bavota}.} \bibinfo{year}{2021}\natexlab{}.
\newblock \showarticletitle{An Empirical Study on the Usage of {BERT} Models
  for Code Completion}. In \bibinfo{booktitle}{\emph{18th {IEEE/ACM}
  International Conference on Mining Software Repositories, {MSR} 2021, Madrid,
  Spain, May 17-19, 2021}}. \bibinfo{publisher}{{IEEE}},
  \bibinfo{pages}{108--119}.
\newblock
\urldef\tempurl%
\url{https://doi.org/10.1109/MSR52588.2021.00024}
\showDOI{\tempurl}


\bibitem[Cogo et~al\mbox{.}(2021)]%
        {cogo2021empirical}
\bibfield{author}{\bibinfo{person}{Filipe~R Cogo}, \bibinfo{person}{Gustavo~A
  Oliva}, \bibinfo{person}{Cor-Paul Bezemer}, {and} \bibinfo{person}{Ahmed~E
  Hassan}.} \bibinfo{year}{2021}\natexlab{}.
\newblock \showarticletitle{An empirical study of same-day releases of popular
  packages in the npm ecosystem}.
\newblock \bibinfo{journal}{\emph{Empirical Software Engineering}}
  \bibinfo{volume}{26}, \bibinfo{number}{5} (\bibinfo{year}{2021}),
  \bibinfo{pages}{1--42}.
\newblock


\bibitem[Cutler et~al\mbox{.}(2012)]%
        {cutler2012random}
\bibfield{author}{\bibinfo{person}{Adele Cutler}, \bibinfo{person}{D~Richard
  Cutler}, {and} \bibinfo{person}{John~R Stevens}.}
  \bibinfo{year}{2012}\natexlab{}.
\newblock \showarticletitle{Random forests}.
\newblock In \bibinfo{booktitle}{\emph{Ensemble machine learning}}.
  \bibinfo{publisher}{Springer}, \bibinfo{pages}{157--175}.
\newblock


\bibitem[Dabbish et~al\mbox{.}(2012)]%
        {dabbish2012social}
\bibfield{author}{\bibinfo{person}{Laura Dabbish}, \bibinfo{person}{Colleen
  Stuart}, \bibinfo{person}{Jason Tsay}, {and} \bibinfo{person}{Jim Herbsleb}.}
  \bibinfo{year}{2012}\natexlab{}.
\newblock \showarticletitle{Social coding in GitHub: transparency and
  collaboration in an open software repository}. In
  \bibinfo{booktitle}{\emph{Proceedings of the ACM 2012 conference on computer
  supported cooperative work}}. \bibinfo{pages}{1277--1286}.
\newblock


\bibitem[Ehsan et~al\mbox{.}(2020)]%
        {ehsan2020empirical}
\bibfield{author}{\bibinfo{person}{Osama Ehsan}, \bibinfo{person}{Safwat
  Hassan}, \bibinfo{person}{Mariam~El Mezouar}, {and} \bibinfo{person}{Ying
  Zou}.} \bibinfo{year}{2020}\natexlab{}.
\newblock \showarticletitle{An empirical study of developer discussions in the
  gitter platform}.
\newblock \bibinfo{journal}{\emph{ACM Transactions on Software Engineering and
  Methodology (TOSEM)}} \bibinfo{volume}{30}, \bibinfo{number}{1}
  (\bibinfo{year}{2020}), \bibinfo{pages}{1--39}.
\newblock


\bibitem[Fan et~al\mbox{.}(2019)]%
        {fan2019impact}
\bibfield{author}{\bibinfo{person}{Yuanrui Fan}, \bibinfo{person}{Xin Xia},
  \bibinfo{person}{Daniel~A COSTA}, \bibinfo{person}{David Lo},
  \bibinfo{person}{Ahmed~E Hassan}, {and} \bibinfo{person}{Shanping Li}.}
  \bibinfo{year}{2019}\natexlab{}.
\newblock \showarticletitle{The impact of changes mislabeled by szz on
  just-in-time defect prediction}.
\newblock \bibinfo{journal}{\emph{IEEE transactions on software engineering}}
  (\bibinfo{year}{2019}), \bibinfo{pages}{1}.
\newblock


\bibitem[Fan et~al\mbox{.}(2018a)]%
        {fan2018chaff}
\bibfield{author}{\bibinfo{person}{Yuanrui Fan}, \bibinfo{person}{Xin Xia},
  \bibinfo{person}{David Lo}, {and} \bibinfo{person}{Ahmed~E Hassan}.}
  \bibinfo{year}{2018}\natexlab{a}.
\newblock \showarticletitle{Chaff from the wheat: Characterizing and
  determining valid bug reports}.
\newblock \bibinfo{journal}{\emph{IEEE transactions on software engineering}}
  \bibinfo{volume}{46}, \bibinfo{number}{5} (\bibinfo{year}{2018}),
  \bibinfo{pages}{495--525}.
\newblock


\bibitem[Fan et~al\mbox{.}(2021)]%
        {fan2021makes}
\bibfield{author}{\bibinfo{person}{Yuanrui Fan}, \bibinfo{person}{Xin Xia},
  \bibinfo{person}{David Lo}, \bibinfo{person}{Ahmed~E Hassan}, {and}
  \bibinfo{person}{Shanping Li}.} \bibinfo{year}{2021}\natexlab{}.
\newblock \showarticletitle{What makes a popular academic AI repository?}
\newblock \bibinfo{journal}{\emph{Empirical Software Engineering}}
  \bibinfo{volume}{26}, \bibinfo{number}{1} (\bibinfo{year}{2021}),
  \bibinfo{pages}{1--35}.
\newblock


\bibitem[Fan et~al\mbox{.}(2018b)]%
        {fan2018early}
\bibfield{author}{\bibinfo{person}{Yuanrui Fan}, \bibinfo{person}{Xin Xia},
  \bibinfo{person}{David Lo}, {and} \bibinfo{person}{Shanping Li}.}
  \bibinfo{year}{2018}\natexlab{b}.
\newblock \showarticletitle{Early prediction of merged code changes to
  prioritize reviewing tasks}.
\newblock \bibinfo{journal}{\emph{Empirical Software Engineering}}
  \bibinfo{volume}{23}, \bibinfo{number}{6} (\bibinfo{year}{2018}),
  \bibinfo{pages}{3346--3393}.
\newblock


\bibitem[Gharehyazie et~al\mbox{.}(2019)]%
        {gharehyazie2019cross}
\bibfield{author}{\bibinfo{person}{Mohammad Gharehyazie},
  \bibinfo{person}{Baishakhi Ray}, \bibinfo{person}{Mehdi Keshani},
  \bibinfo{person}{Masoumeh~Soleimani Zavosht}, \bibinfo{person}{Abbas
  Heydarnoori}, {and} \bibinfo{person}{Vladimir Filkov}.}
  \bibinfo{year}{2019}\natexlab{}.
\newblock \showarticletitle{Cross-project code clones in GitHub}.
\newblock \bibinfo{journal}{\emph{Empirical Software Engineering}}
  \bibinfo{volume}{24}, \bibinfo{number}{3} (\bibinfo{year}{2019}),
  \bibinfo{pages}{1538--1573}.
\newblock


\bibitem[Hearst et~al\mbox{.}(1998)]%
        {hearst1998svm}
\bibfield{author}{\bibinfo{person}{M.A. Hearst}, \bibinfo{person}{S.T. Dumais},
  \bibinfo{person}{E. Osuna}, \bibinfo{person}{J. Platt}, {and}
  \bibinfo{person}{B. Scholkopf}.} \bibinfo{year}{1998}\natexlab{}.
\newblock \showarticletitle{Support vector machines}.
\newblock \bibinfo{journal}{\emph{IEEE Intelligent Systems and their
  Applications}} \bibinfo{volume}{13}, \bibinfo{number}{4}
  (\bibinfo{year}{1998}), \bibinfo{pages}{18--28}.
\newblock
\urldef\tempurl%
\url{https://doi.org/10.1109/5254.708428}
\showDOI{\tempurl}


\bibitem[Jebnoun et~al\mbox{.}(2020)]%
        {jebnoun2020scent}
\bibfield{author}{\bibinfo{person}{Hadhemi Jebnoun}, \bibinfo{person}{Houssem
  Ben~Braiek}, \bibinfo{person}{Mohammad~Masudur Rahman}, {and}
  \bibinfo{person}{Foutse Khomh}.} \bibinfo{year}{2020}\natexlab{}.
\newblock \showarticletitle{The scent of deep learning code: An empirical
  study}. In \bibinfo{booktitle}{\emph{Proceedings of the 17th International
  Conference on Mining Software Repositories}}. \bibinfo{pages}{420--430}.
\newblock


\bibitem[Joblin and Apel(2021)]%
        {mitchell2022}
\bibfield{author}{\bibinfo{person}{Mitchell Joblin} {and} \bibinfo{person}{Sven
  Apel}.} \bibinfo{year}{2021}\natexlab{}.
\newblock \showarticletitle{How Do Successful and Failed Projects Differ? A
  Socio-Technical Analysis}.
\newblock \bibinfo{journal}{\emph{ACM Trans. Softw. Eng. Methodol.}}
  (\bibinfo{date}{dec} \bibinfo{year}{2021}).
\newblock
\showISSN{1049-331X}
\urldef\tempurl%
\url{https://doi.org/10.1145/3504003}
\showDOI{\tempurl}


\bibitem[Kavaler et~al\mbox{.}(2017)]%
        {kavaler2017perceived}
\bibfield{author}{\bibinfo{person}{David Kavaler}, \bibinfo{person}{Sasha
  Sirovica}, \bibinfo{person}{Vincent Hellendoorn}, \bibinfo{person}{Raul
  Aranovich}, {and} \bibinfo{person}{Vladimir Filkov}.}
  \bibinfo{year}{2017}\natexlab{}.
\newblock \showarticletitle{Perceived language complexity in GitHub issue
  discussions and their effect on issue resolution}. In
  \bibinfo{booktitle}{\emph{2017 32nd IEEE/ACM International Conference on
  Automated Software Engineering (ASE)}}. IEEE, \bibinfo{pages}{72--83}.
\newblock


\bibitem[Lamba et~al\mbox{.}(2020)]%
        {lamba2020heard}
\bibfield{author}{\bibinfo{person}{Hemank Lamba}, \bibinfo{person}{Asher
  Trockman}, \bibinfo{person}{Daniel Armanios}, \bibinfo{person}{Christian
  K{\"a}stner}, \bibinfo{person}{Heather Miller}, {and} \bibinfo{person}{Bogdan
  Vasilescu}.} \bibinfo{year}{2020}\natexlab{}.
\newblock \showarticletitle{Heard it through the Gitvine: an empirical study of
  tool diffusion across the npm ecosystem}. In
  \bibinfo{booktitle}{\emph{Proceedings of the 28th ACM Joint Meeting on
  European Software Engineering Conference and Symposium on the Foundations of
  Software Engineering}}. \bibinfo{pages}{505--517}.
\newblock


\bibitem[Lee et~al\mbox{.}(2013)]%
        {lee2013github}
\bibfield{author}{\bibinfo{person}{Michael~J Lee}, \bibinfo{person}{Bruce
  Ferwerda}, \bibinfo{person}{Junghong Choi}, \bibinfo{person}{Jungpil Hahn},
  \bibinfo{person}{Jae~Yun Moon}, {and} \bibinfo{person}{Jinwoo Kim}.}
  \bibinfo{year}{2013}\natexlab{}.
\newblock \showarticletitle{GitHub developers use rockstars to overcome
  overflow of news}.
\newblock In \bibinfo{booktitle}{\emph{CHI'13 Extended Abstracts on Human
  Factors in Computing Systems}}. \bibinfo{pages}{133--138}.
\newblock


\bibitem[Liu et~al\mbox{.}(2021)]%
        {liu2021reproducibility}
\bibfield{author}{\bibinfo{person}{Chao Liu}, \bibinfo{person}{Cuiyun Gao},
  \bibinfo{person}{Xin Xia}, \bibinfo{person}{David Lo}, \bibinfo{person}{John
  Grundy}, {and} \bibinfo{person}{Xiaohu Yang}.}
  \bibinfo{year}{2021}\natexlab{}.
\newblock \showarticletitle{On the Reproducibility and Replicability of Deep
  Learning in Software Engineering}.
\newblock \bibinfo{journal}{\emph{ACM Transactions on Software Engineering and
  Methodology (TOSEM)}} \bibinfo{volume}{31}, \bibinfo{number}{1}
  (\bibinfo{year}{2021}), \bibinfo{pages}{1--46}.
\newblock


\bibitem[Liu et~al\mbox{.}(2022)]%
        {liu2022readme}
\bibfield{author}{\bibinfo{person}{Yuyang Liu}, \bibinfo{person}{Ehsan Noei},
  {and} \bibinfo{person}{Kelly Lyons}.} \bibinfo{year}{2022}\natexlab{}.
\newblock \showarticletitle{How ReadMe files are structured in open source Java
  projects}.
\newblock \bibinfo{journal}{\emph{Information and Software Technology}}
  \bibinfo{volume}{148} (\bibinfo{year}{2022}), \bibinfo{pages}{106924}.
\newblock


\bibitem[Macbeth et~al\mbox{.}(2011)]%
        {macbeth2011cliff}
\bibfield{author}{\bibinfo{person}{Guillermo Macbeth}, \bibinfo{person}{Eugenia
  Razumiejczyk}, {and} \bibinfo{person}{Rub{\'e}n~Daniel Ledesma}.}
  \bibinfo{year}{2011}\natexlab{}.
\newblock \showarticletitle{Cliff's Delta Calculator: A non-parametric effect
  size program for two groups of observations}.
\newblock \bibinfo{journal}{\emph{Universitas Psychologica}}
  \bibinfo{volume}{10}, \bibinfo{number}{2} (\bibinfo{year}{2011}),
  \bibinfo{pages}{545--555}.
\newblock


\bibitem[Milewicz et~al\mbox{.}(2019)]%
        {milewicz2019characterizing}
\bibfield{author}{\bibinfo{person}{Reed Milewicz}, \bibinfo{person}{Gustavo
  Pinto}, {and} \bibinfo{person}{Paige Rodeghero}.}
  \bibinfo{year}{2019}\natexlab{}.
\newblock \showarticletitle{Characterizing the roles of contributors in
  open-source scientific software projects}. In \bibinfo{booktitle}{\emph{2019
  IEEE/ACM 16th International Conference on Mining Software Repositories
  (MSR)}}. IEEE, \bibinfo{pages}{421--432}.
\newblock


\bibitem[Murgia et~al\mbox{.}(2014)]%
        {murgia2014influence}
\bibfield{author}{\bibinfo{person}{Alessandro Murgia}, \bibinfo{person}{Giulio
  Concas}, \bibinfo{person}{Roberto Tonelli}, \bibinfo{person}{Marco Ortu},
  \bibinfo{person}{Serge Demeyer}, {and} \bibinfo{person}{Michele Marchesi}.}
  \bibinfo{year}{2014}\natexlab{}.
\newblock \showarticletitle{On the influence of maintenance activity types on
  the issue resolution time}. In \bibinfo{booktitle}{\emph{Proceedings of the
  10th international conference on predictive models in software engineering}}.
  \bibinfo{pages}{12--21}.
\newblock


\bibitem[Overney(2020)]%
        {overney2020hanging}
\bibfield{author}{\bibinfo{person}{Cassandra Overney}.}
  \bibinfo{year}{2020}\natexlab{}.
\newblock \showarticletitle{Hanging by the thread: an empirical study of
  donations in open source}. In \bibinfo{booktitle}{\emph{Proceedings of the
  ACM/IEEE 42nd International Conference on Software Engineering: Companion
  Proceedings}}. \bibinfo{pages}{131--133}.
\newblock


\bibitem[Prana et~al\mbox{.}(2021)]%
        {prana2021including}
\bibfield{author}{\bibinfo{person}{Gede Artha~Azriadi Prana},
  \bibinfo{person}{Denae Ford}, \bibinfo{person}{Ayushi Rastogi},
  \bibinfo{person}{David Lo}, \bibinfo{person}{Rahul Purandare}, {and}
  \bibinfo{person}{Nachiappan Nagappan}.} \bibinfo{year}{2021}\natexlab{}.
\newblock \showarticletitle{Including everyone, everywhere: Understanding
  opportunities and challenges of geographic gender-inclusion in oss}.
\newblock \bibinfo{journal}{\emph{IEEE Transactions on Software Engineering}}
  (\bibinfo{year}{2021}).
\newblock


\bibitem[Prana et~al\mbox{.}(2019)]%
        {prana2019categorizing}
\bibfield{author}{\bibinfo{person}{Gede Artha~Azriadi Prana},
  \bibinfo{person}{Christoph Treude}, \bibinfo{person}{Ferdian Thung},
  \bibinfo{person}{Thushari Atapattu}, {and} \bibinfo{person}{David Lo}.}
  \bibinfo{year}{2019}\natexlab{}.
\newblock \showarticletitle{Categorizing the content of github readme files}.
\newblock \bibinfo{journal}{\emph{Empirical Software Engineering}}
  \bibinfo{volume}{24}, \bibinfo{number}{3} (\bibinfo{year}{2019}),
  \bibinfo{pages}{1296--1327}.
\newblock


\bibitem[Rahman and Farhana(2021)]%
        {rahman2021empirical}
\bibfield{author}{\bibinfo{person}{Akond Rahman} {and} \bibinfo{person}{Effat
  Farhana}.} \bibinfo{year}{2021}\natexlab{}.
\newblock \showarticletitle{An Empirical Study of Bugs in COVID19 Software
  Projects}.
\newblock \bibinfo{journal}{\emph{Journal of Software Engineering}}
  \bibinfo{volume}{9} (\bibinfo{year}{2021}), \bibinfo{pages}{3}.
\newblock


\bibitem[Treude et~al\mbox{.}(2020)]%
        {treude2020beyond}
\bibfield{author}{\bibinfo{person}{Christoph Treude}, \bibinfo{person}{Justin
  Middleton}, {and} \bibinfo{person}{Thushari Atapattu}.}
  \bibinfo{year}{2020}\natexlab{}.
\newblock \showarticletitle{Beyond accuracy: Assessing software documentation
  quality}. In \bibinfo{booktitle}{\emph{Proceedings of the 28th ACM Joint
  Meeting on European Software Engineering Conference and Symposium on the
  Foundations of Software Engineering}}. \bibinfo{pages}{1509--1512}.
\newblock


\bibitem[Trockman et~al\mbox{.}(2018)]%
        {trockman2018adding}
\bibfield{author}{\bibinfo{person}{Asher Trockman}, \bibinfo{person}{Shurui
  Zhou}, \bibinfo{person}{Christian K{\"a}stner}, {and} \bibinfo{person}{Bogdan
  Vasilescu}.} \bibinfo{year}{2018}\natexlab{}.
\newblock \showarticletitle{Adding sparkle to social coding: an empirical study
  of repository badges in the npm ecosystem}. In
  \bibinfo{booktitle}{\emph{Proceedings of the 40th International Conference on
  Software Engineering}}. \bibinfo{pages}{511--522}.
\newblock


\bibitem[Tsay et~al\mbox{.}(2014)]%
        {tsay2014let}
\bibfield{author}{\bibinfo{person}{Jason Tsay}, \bibinfo{person}{Laura
  Dabbish}, {and} \bibinfo{person}{James Herbsleb}.}
  \bibinfo{year}{2014}\natexlab{}.
\newblock \showarticletitle{Let's talk about it: evaluating contributions
  through discussion in GitHub}. In \bibinfo{booktitle}{\emph{Proceedings of
  the 22nd ACM SIGSOFT international symposium on foundations of software
  engineering}}. \bibinfo{pages}{144--154}.
\newblock


\bibitem[Verdi et~al\mbox{.}(2020)]%
        {verdi2020empirical}
\bibfield{author}{\bibinfo{person}{Morteza Verdi}, \bibinfo{person}{Ashkan
  Sami}, \bibinfo{person}{Jafar Akhondali}, \bibinfo{person}{Foutse Khomh},
  \bibinfo{person}{Gias Uddin}, {and} \bibinfo{person}{Alireza~Karami
  Motlagh}.} \bibinfo{year}{2020}\natexlab{}.
\newblock \showarticletitle{An empirical study of c++ vulnerabilities in
  crowd-sourced code examples}.
\newblock \bibinfo{journal}{\emph{IEEE Transactions on Software Engineering}}
  (\bibinfo{year}{2020}).
\newblock


\bibitem[Yu et~al\mbox{.}(2015)]%
        {yu2015wait}
\bibfield{author}{\bibinfo{person}{Yue Yu}, \bibinfo{person}{Huaimin Wang},
  \bibinfo{person}{Vladimir Filkov}, \bibinfo{person}{Premkumar Devanbu}, {and}
  \bibinfo{person}{Bogdan Vasilescu}.} \bibinfo{year}{2015}\natexlab{}.
\newblock \showarticletitle{Wait for it: Determinants of pull request
  evaluation latency on github}. In \bibinfo{booktitle}{\emph{2015 IEEE/ACM
  12th working conference on mining software repositories}}. IEEE,
  \bibinfo{pages}{367--371}.
\newblock


\bibitem[Zhang et~al\mbox{.}(2017)]%
        {zhang2017detecting}
\bibfield{author}{\bibinfo{person}{Yun Zhang}, \bibinfo{person}{David Lo},
  \bibinfo{person}{Pavneet~Singh Kochhar}, \bibinfo{person}{Xin Xia},
  \bibinfo{person}{Quanlai Li}, {and} \bibinfo{person}{Jianling Sun}.}
  \bibinfo{year}{2017}\natexlab{}.
\newblock \showarticletitle{Detecting similar repositories on GitHub}. In
  \bibinfo{booktitle}{\emph{2017 IEEE 24th International Conference on Software
  Analysis, Evolution and Reengineering (SANER)}}. IEEE,
  \bibinfo{pages}{13--23}.
\newblock


\bibitem[Zhu et~al\mbox{.}(2014)]%
        {zhu2014patterns}
\bibfield{author}{\bibinfo{person}{Jiaxin Zhu}, \bibinfo{person}{Minghui Zhou},
  {and} \bibinfo{person}{Audris Mockus}.} \bibinfo{year}{2014}\natexlab{}.
\newblock \showarticletitle{Patterns of folder use and project popularity: A
  case study of GitHub repositories}. In \bibinfo{booktitle}{\emph{Proceedings
  of the 8th ACM/IEEE International Symposium on Empirical Software Engineering
  and Measurement}}. \bibinfo{pages}{1--4}.
\newblock


\end{thebibliography}

\end{document}